\documentclass[authoryear,3p,review,12pt,hidelinks]{elsarticle}

\usepackage{amsmath}
\usepackage{amssymb}
\usepackage{booktabs}
\usepackage{float}
\usepackage{orcidlink}
\usepackage{lineno}

\journal{Expert Systems with Applications}

\begin{document}

\begin{frontmatter}



\title{Neural Networks for Fast Optimisation in Model Predictive Control: A Review}


\author[iisri]{Camilo Gonzalez Arango\corref{cor1}\,\orcidlink{0000-0002-3874-4801}}
\ead{c.gonzalezarango@deakin.edu.au}

\author[iisri]{Houshyar Asadi\,\orcidlink{0000-0002-3620-8693}}
\ead{houshyar.asadi@deakin.edu.au}

\author[impact]{Lars Kooijman\,\orcidlink{0000-0002-0902-5752}}
\ead{lars.kooijman@monash.edu}

\author[swin]{Chee Peng Lim}
\ead{cplim@swin.edu.au}

\affiliation[iisri]{organization={Deakin University, Institute for Intelligent Systems Research and Innovation},
            addressline={75 Pigdons Rd}, 
            city={Waurn Ponds},
            postcode={3216}, 
            state={Victoria},
            country={Australia}}
\affiliation[impact]{organization={Monash University, Opportunity Tech Lab},
            addressline={900 Dandenong Rd}, 
            city={Caulfield East},
            postcode={3145}, 
            state={Victoria},
            country={Australia}}
\affiliation[swin]{organization={Swinburne University of Technology, Swinburne Research},
            addressline={John St}, 
            city={Hawthorn},
            postcode={3122}, 
            state={Victoria},
            country={Australia}}
\cortext[cor1]{Corresponding author.}

\begin{abstract}
Model Predictive Control (MPC) is an optimal control algorithm with strong stability and robustness guarantees. Despite its popularity in robotics and industrial applications, the main challenge in deploying MPC is its high computation cost, stemming from the need to solve an optimisation problem at each control interval. There are several methods to reduce this cost. This survey focusses on approaches where a neural network is used to approximate an existing controller. Herein, relevant and unique neural approximation methods for linear, nonlinear, and robust MPC are presented and compared. Comparisons are based on the theoretical guarantees that are preserved, the factor by which the original controller is sped up, and the size of problem that a framework is applicable to. Research contributions include: a taxonomy that organises existing knowledge, a summary of literary gaps, discussion on promising research directions, and simple guidelines for choosing an approximation framework. The main conclusions are that (1) new benchmarking tools are needed to help prove the generalisability and scalability of approximation frameworks, (2) future breakthroughs most likely lie in the development of ties between control and learning, and (3) the potential and applicability of recently developed neural architectures and tools remains unexplored in this field.
\end{abstract}


\end{frontmatter}


\section{Introduction}
\label{sec:Intro}

Model Predictive Control (MPC) is an optimal control scheme that is widely used in the process industry and robotics. Compared to other control schemes, its main advantages include seamless handling of MIMO systems, ease of consideration of state, input and output constraints, the guarantee of closed loop stability by design, and the ability to exploit knowledge and predictions of future control references or disturbances \citep{schwenzer2021review, kouvaritakis2015model}. In a nutshell, MPC controllers use a model or ``plant" of the controlled system to derive a sequence of control inputs that will result in optimal system response over a future period of time, or ``horizon", given the current states of the system and desired reference trajectories. This is achieved by solving an optimisation problem where the design variables are the sequence of future system inputs and the plant is used to calculate future system outputs and states for said sequence. The primary challenge in deploying MPC controllers in practical applications stems from the necessity to solve large optimisation problems in real time. For systems with fast dynamics requiring control at high frequency, this control scheme becomes infeasible even for simple systems, especially when the more complex formulations of MPC such as Nonlinear MPC (NMPC) or Robust MPC (RMPC) are used, or when the controller needs to be deployed in low cost embedded platforms with limited computing resources. 

\begin{figure}[!b]
    \centering
    \includegraphics[width=\linewidth]{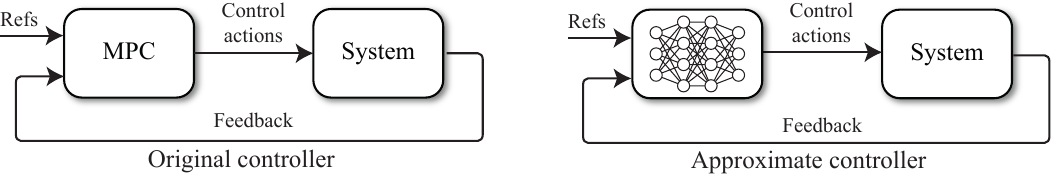}
    \caption{Typical implementation of NNBO to replace an MPC controller.}
    \label{fig:introFigure}
\end{figure}

Many approaches have been proposed to deal with the computational cost of MPC and facilitate its use in real-world applications. These approaches can be broadly categorised as efficient optimisation solvers, Explicit MPC (EMPC), and approximate optimisation methods. Efficient solvers are specialised optimisation routines designed for specific problem formulations; they generally make use of hot start techniques and early stopping to reduce computation times. This type of approach can succeed in enabling some MPC implementations in real time but for complex forms of MPC or where long horizons are considered, the reduction in computation time is often insufficient. For more information on such solvers, refer to the review by \cite{ferreau2017embedded}. Alternatively, EMPC is a form of MPC where the optimisation problem is solved offline and stored as lookup tables that map system states and target outputs to optimal control inputs. This can be done exactly for Linear MPC (LMPC) \citep{bemporad2002explicit} and approximately for NMPC \citep{johansen2004approximate, bemporad2006algorithm}. However, producing such lookup tables requires solving large multi-parametric programs which often become intractable for large systems, and when solutions are possible, lookup tables grow exponentially with horizon length to sizes that do not fit in the limited memory of embedded controllers, or that are still too slow to query in real time \citep{karg2020efficient, fabiani2022reliably}. For a survey of EMPC refer to \cite{alessio2009survey}. Lastly, approximate optimisation methods are approaches where conventional solvers are replaced with alternatives that can produce similar outputs for the same inputs but are much cheaper to evaluate in terms of computation cost. A particularly effective way of doing this is to replace the optimisation solver by a Neural Network (NN) that has been trained to imitate the solver thereby reducing the optimisation operation to a simple function evaluation, Figure \ref{fig:introFigure} illustrates how this idea can be applied to MPC. We refer to this type of approach as Neural Network-Based Optimisation (NNBO), neural solvers, or neural approximations.
Compared to efficient solvers and EMPC, NNBO can be applied to all types of MPC, it requires insignificant storage space, and the reduction in computation time can be several orders of magnitude. Therefore, it is often the only viable option for real-time deployment of many MPC controllers. Nonetheless, NNBO has challenges of its own, such as the loss of rigorous constraint satisfaction, stability, and recursive feasibility guarantees compared to exact MPC, as well as, the unavoidable regression error between the NN outputs and those of the real solver.

The use, benefits, and limitations of efficient optimisation solvers and EMPC have been presented and discussed on previous reviews \citep{ferreau2017embedded, alessio2009survey}, yet a review on the use of NNs as a replacement of the solver in MPC remains absent in literature. \cite{norouzi2023integrating} touched on this topic in their review of ML in MPC for automotive applications. However, the discussion on the matter was limited to four studies which do not encompass all of the existing approximation frameworks. Similarly, \cite{hewing2020learning} and \cite{mesbah2022fusion} briefly discussed the subject in their reviews on ways in which ML has been integrated in MPC. But again, the scope of those works is broad and does not go into detail on the subject of NNBO. Herein, a survey is presented of the various frameworks and methods that can be used to generate approximate MPC controllers powered by NNBO. Compared to the aforementioned works \citep{norouzi2023integrating, hewing2020learning, mesbah2022fusion}, we focus exclusively on this topic. Moreover, we present and compare a wide range of unique existing frameworks including discussions on how the different approaches deal with the challenges of NNBO. Our contributions can be summarised as follows: 
\begin{itemize}
    \item We provide a critical review of methods to generate NN approximations of MPC including detailed discussion of the advantages and disadvantages of the different methods.
    \item We propose a taxonomy that indicates which challenges a certain method is relevant to, and how the challenge is addressed. Future studies can utilise this taxonomy to label their methods in a traceable manner that precisely conveys the capabilities of the method.
    \item  We provide a selection guide synthesised from case studies which indicates which methods are relevant to which types of MPC, plant size, and horizon length. Given that most approaches have only been demonstrated in simulation, we emphasise methods that have been proven to be effective in real systems.
    \item We provide a summary of the gaps that remain unaddressed in the literature on NNBO for MPC and promising avenues for future work. 
\end{itemize}

The remainder of the paper is organised as follows, Section \ref{sec:RelatedWork} explains the scope of the survey and the type of method that it covers with reference to related surveys that cover out-of-scope topics. Section \ref{sec:Background} provides a general introduction to MPC and introduces our proposed taxonomy of NNBO methods. In Sections \ref{sec:NNBO4LMPC} -- \ref{sec:NNBO4RMPC}, we introduce and explain different frameworks that can be used to approximate LMPC, NMPC, and RMPC with NNBO, respectively. We compare the different methods according to the type of guarantees they offer, the size of problem they have been validated on, the factor by which they speed up the controller, and evidence showcasing their success beyond simulation. 

\section{Scope and Related Work}
\label{sec:RelatedWork}
Machine Learning (ML) techniques have been extensively used in the field of MPC for various purposes besides accelerating optimisation, therefore, the scope of this survey excludes certain topics. First of all, we differentiate between approaches where the MPC optimiser is replaced with an NN that requires a single evaluation (e.g. Feedforward NNs), and approaches stemming from the work by \cite{tank1986simple}, where the optimiser is replaced by a Recurrent NN (RNN) that needs to be simulated iteratively until convergence to yield a prediction. This survey focuses on approaches that necessitate a single evaluation. Exhaustive surveys on RNNs for optimisation, often referred to as Hopfield NNs, have been presented by \cite{wen2009review} and \cite{jin2019survey}. Furthermore, we omit publications that utilise ML to construct data-driven models for use in MPC, unless these methods are integrated within a framework that concurrently approximates and expedites the optimiser. Although data-driven models can lead to reductions in computation time, their purpose is to achieve better plant predictive accuracy, not sufficient speed for implementation in real time. The works by \cite{ren2022tutorial, norouzi2023integrating, mesbah2022fusion, hewing2020learning} and \cite{meng2022emerging} provide good coverage of such approaches. In addition, we do not cover the topic of using ML techniques, or most commonly Reinforcement Learning (RL), to find parametrisations of the cost function in MPC that yield better performance or enable online controller adaptation. Approaches of this nature have been assessed in previous reviews \citep{mesbah2022fusion, hewing2020learning} and are designed to enhance controllers that are already suitable for real-time deployment, rather than facilitating their initial deployment. However, where relevant we do discuss the subject of improving neural approximations of MPC with RL. Similarly, we only briefly touch on the topic of initialising RL agents with behaviours learnt from MPC \citep{karg2021reinforced}. Such approaches are an extension of the methods reviewed here and a complete review is outside the scope of the paper. We also note that there are several publications (e.g. \citep{chakrabarty2016support, bemporad2011ultra, csekHo2015explicit, karg2020efficient}) on the topic of learning EMPC policies to reduce the computational burden of the lookup operations at run time. We exclude such approaches from our scope because they rely on being able to generate an EMPC controller in the first place, which is not always possible. The methods discussed here are not constrained by this limitation. Lastly, we only provide a high level introduction to MPC and the different variants that exist. For an in depth dive into the theory of MPC, readers are directed \citep{kouvaritakis2015model}. Those seeking an application-oriented perspective are directed to \cite{schwenzer2021review} and \cite{rakovic2018handbook}.

\section{Background}
\label{sec:Background}
\subsection{Model Predictive Control Basics}
\begin{figure}[!t]
    \centering
    \includegraphics[width=\linewidth]{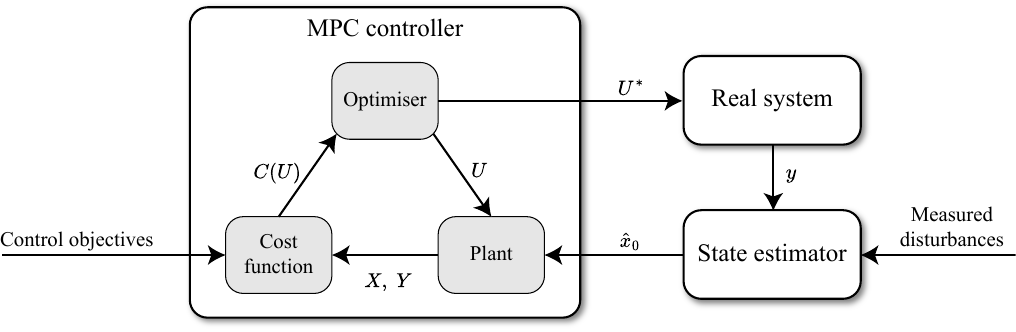}
    \caption{Block diagram of typical implementation and components of MPC solution.}
    \label{fig:MPCArch}
\end{figure}

An MPC controller implementation typically requires three components: an optimisation solver, a plant model, and a state estimator, as shown in the block diagram in Figure \ref{fig:MPCArch}. The optimisation solver is one of many existing algorithms chosen according to the type of problem involved and it is used to solve the underlying optimisation problem in MPC. Commonly encountered problem types in MPC are Quadratic Programs (QPs), Nonlinear Programming (NLP) problems, min-max problems, and Mixed Integer Linear Programs (MILPs) \citep{ferreau2017embedded, kouvaritakis2015model}. The plant model is a function that captures the dynamics of the system being controlled and enables the simulation of the system responses to inputs. Lastly, the state estimator is an algorithm that can be used to approximate system states and disturbances that cannot be directly measured from other variables that can be measured. The most popular realisation of this subsystem is a Kalman filter \citep{rakovic2018handbook} but it should be noted that a state estimator is not always required; in some systems all variables are directly measurable. From hereon, we focus the discussion on the optimiser and plant blocks.
The generalised form of a discrete-time plant model is given by \eqref{eq:generalDTPlant}, where the subscript $i$ indicates the current time step, $t$ is the current time, and $x \in \mathbb{R}^{n_x}$, $u \in \mathbb{R}^{n_u}$, and $y \in \mathbb{R}^{n_y}$ are the state, input and output vectors of the system \citep{hewing2020learning}. The uncertainty of the system is characterised by the variables $\theta$, $w$, and $v$. More specifically, $\theta$ accounts for parametric model uncertainties and is constant over time, $w$ accounts for disturbances and process noise, and $v$ accounts for output measurement noise. Both $w$ and $v$ are made up of independent and identically distributed (i.i.d). time varying random variables. Lastly, the functions $f$ and $g$ are the state and output transition functions, which may take any arbitrary form, for instance a linear state space model, a neural network, a lookup table, etc. 
\begin{subequations}\label{eq:generalDTPlant}
\begin{align}
  x_{i+1} = f(x_i, u_i, t, w_i, \theta) \tag{\ref{eq:generalDTPlant}} \\
  y_i = g(x_i, u_i, t, v_i, \theta) \label{eq:generalDTPlant_output}
\end{align}
\end{subequations}

At each control interval the optimiser uses the plant model to solve an optimisation problem of the form shown in \eqref{eq:generalCostFun}
\begin{subequations}\label{eq:generalCostFun}
\begin{align}
  \min_{U} \quad & C(U) = \sum_{i=0}^{N_p-1} l(x_i, u_i, i) + l_f(x_{N_P}, u_{N_P}, N_p)  \tag{\ref{eq:generalCostFun}} \\ 
  \textrm{s.t.} \quad & x_{i+1} = f(x_i, u_i, t, w_i, \theta)  \label{eq:generalCostFun_a} \\
                    & y_i = g(x_i, u_i, t, v_i, \theta)  \label{eq:generalCostFun_b} \\
                    & U = [u_0, \ldots, u_N] \in \mathcal{U}_j \; \forall j=1,\ldots, n_u  \label{eq:generalCostFun_c} \\
                    & Y = [y_0, \ldots, y_N] \in \mathcal{Y}_j \; \forall j=1,\ldots, n_y  \label{eq:generalCostFun_d} \\
                    & X = [x_0, \ldots, x_N] \in \mathcal{X}_j \; \forall j=1,\ldots, n_x  \label{eq:generalCostFun_e} \\
                    & x_N \in \mathcal{X}_f \label{eq:generalCostFun_f}
\end{align}
\end{subequations}
where $l$ and $l_f$ are the stage and terminal cost functions, $N_p$ is the prediction horizon length (i.e. the number of time steps that the controller looks forward into the future), $\mathcal{U}_j$, $\mathcal{Y}_j$ and $\mathcal{X}_j$ are generalised forms of equality and inequality constraints expressed as sets, and $\mathcal{X}_f$ is the terminal states set. The cost function $l$ is typically designed to achieve control objectives such as reference tracking, trajectory tracking, path tracking, control input smoothing, and control input minimisation. For each state, output, and input one may choose to use none, one, or several of these objectives depending on the desired behaviour. The equality and inequality constraints \eqref{eq:generalCostFun_a} - \eqref{eq:generalCostFun_e} are used to enforce the system dynamics and all input, output and state constraints over the horizon. Lastly, the constraint \eqref{eq:generalCostFun_f} and the cost $l_f$ are used to enforce terminal conditions, which, in certain types of MPC, provide assurances of stability and recursive feasibility. The solution of \eqref{eq:generalCostFun} is the optimal sequence of control inputs ($U^*$) to be applied to the system which will drive it to its desired states over the prediction horizon. In practice, only the first element of $U^*$ is applied to the system and the entire control sequence is recomputed at the next control interval by solving the optimisation problem again. It should be noted that many MPC implementations also distinguish between the prediction horizon and the control horizon, which is a shorter window of future time of length ($N_c$) over which control actions are optimised, that is ($N_c < N_p$) \citep{rakovic2018handbook}.

In the following subsections, we provide a short description of the different types of MPC, including detail about the specific forms of \eqref{eq:generalDTPlant} and \eqref{eq:generalCostFun} that are used. This information is important to understand the benefits and challenges of replacing the optimiser block in Figure \ref{fig:MPCArch} with an NNBO method.

\subsubsection{Linear Model Predictive Control}

Linear MPC (LMPC) is the most basic form of MPC. It is limited to employing linear plant models alongside a quadratic cost function coupled with linear constraints. The typical choice of plant is a time invariant state space model of the form shown in \eqref{eq:StandardStateSpace} \citep{kouvaritakis2015model}. However, it is also feasible to utilise time-varying plants. For time invariant plants the matrices $A$, $B$, $C$, and $D$ in \eqref{eq:StandardStateSpace} are fixed, whereas for time-varying plants these matrices can change over time (e.g. $A = A(t)$). In practice, time-varying plants can be used to linearise complex systems at various operating points, thus offering more accurate modelling. Due to this distinction, we refer to LMPC that uses a time-varying plant as Linear Time Varying MPC (LVMPC). 
\begin{align}
    \label{eq:StandardStateSpace}
    x_{i+1} &= A x_i + B u_i \\
    y_i & = C x_i + D u_i \nonumber
\end{align}
Both LMPC and LVMPC use the same form of cost function, an example of which is shown in \eqref{eq:GeneralMPCFormulation}. The first term encourages reference output $(y_r)$ tracking and the second term encourages reference input $(u_r)$ tracking and control input smoothing. The states, outputs and inputs can be subject to polytopic bounds, and $\bar{x}$ is the initial state of the system. For MIMO systems, the weight matrices $\Lambda_y$, $\Lambda_u$ and $\Lambda_{du}$ control the relative importance of the various references.
\begin{align}
    \label{eq:GeneralMPCFormulation}
    \min_{U} \quad & \sum_{i=0}^{N_p} (y_i-y_r)^\intercal \Lambda_y (y_i-y_r)
                            + \sum_{i=0}^{N_c} (u_i-u_r)^\intercal \Lambda_u (u_i-u_r) + (\Delta u_i - \Delta u_r)^\intercal \Lambda_{du} (\Delta u_i - \Delta u_r)\\ \nonumber
    \textrm{s.t.} \quad & x_{i+1} = A x_i + B u_i \\ \nonumber
                        & x_{\rm min} \le x_i  \le x_{\rm max} \\ \nonumber
                        & y_{\rm min} \le y_i  \le y_{\rm max} \\ \nonumber
                        & u_{\rm min} \le u_i  \le u_{\rm max} \\ \nonumber
                        & x_0 = \bar{x}\nonumber
\end{align}
Through algebraic manipulation \eqref{eq:GeneralMPCFormulation} can be restructured into a standard QP the form of which is shown in \eqref{eq:StandardQP}. For LMPC $P$, $q$, and $R$, are fixed. Whereas, for LVMPC, these matrices vary over time to capture the changing dynamics of the plant. The format in \eqref{eq:StandardQP} enables much more efficient optimisation and many optimisers can solve large versions of this problem in real time at high frequencies, especially if the problem is strictly convex \citep{ferreau2014qpoases, osqp}. 
\begin{align}
    \label{eq:StandardQP}
    \min_{U} \quad & \frac{1}{2}U^\intercal\, P\, U + q^\intercal\, U \\ \nonumber
    \textrm{s.t.} \quad & l_b \le R\, U \le u_b \nonumber
\end{align}

\subsubsection{Nonlinear Model Predictive Control}

NMPC is a more flexible formulation of MPC that allows the use of nonlinear plant models, any form of cost function, and nonlinear constraints \citep{alessio2009survey}. The plant model need not conform to a specific structure (e.g. a state space model), it can be represented in any manner that is most suitable or efficient for addressing the problem at hand. The generalised form of an NMPC plant and optimisation problem is already well captured by Equations \eqref{eq:generalDTPlant} and \eqref{eq:generalCostFun}. The higher flexibility of NMPC comes at the cost of increased difficulty in solving the underlying optimisation problem which is typically of the NLP type. In theory, the plant, cost and constraints can take any arbitrary form, but in practice most implementations restrict these choices to forms that are cheap to evaluate and result in problems that can be solved efficiently, for instance using Sequential Quadratic Programming (SQP). Thus, the capabilities of NMPC are rarely fully exploited in real-world applications, NNBO can help address this limitation. 

\subsubsection{Robust Model Predictive Control}

RMPC is designed to generate control actions that account for the possible realisations of uncertainties and disturbances over the prediction horizon. The previously discussed types of MPC optimise control actions based on a single set of plant predictions over the prediction horizon. In contrast, RMPC aims to optimise for a distribution of possible predictions, or for multiple samples drawn from said distribution. The distribution or samples correspond to all or particular ways in which the system may behave under different realisations of uncertainties and disturbances. Considering more than one prediction greatly increases the dimensionality and computational complexity of the underlying optimisation problem. However, there are formulations of RMPC that can do so efficiently by considering the statistical properties of uncertainties and disturbances, or by considering only a few relevant realisations of such values over a short horizon termed the robust horizon ($N_r$). Examples of the former formulations include min-max RMPC \citep{witsenhausen1968minimax, scokaert1998min, diehl2004robust} and Tube RMPC \citep{mayne2005robust, mayne2011tube} while the latter formulation refers to multi stage RMPC (msRMPC) \citep{lucia2013multi, lucia2015robust}, also known as scenario RMPC. The plant models used in these types of MPC are the same as for LMPC and RMPC but contain additional terms and parameters that account for uncertainties and disturbances as in \eqref{eq:generalDTPlant}. Cost functions, on the other hand, are highly formulation-specific. As for NMPC, the computational cost of RMPC is a great motivation for the use of NNBO instead of conventional solvers. 

\subsection{Taxonomy of Surveyed Methods}

\begin{figure}[!b]
    \centering
    \includegraphics[width=11cm]{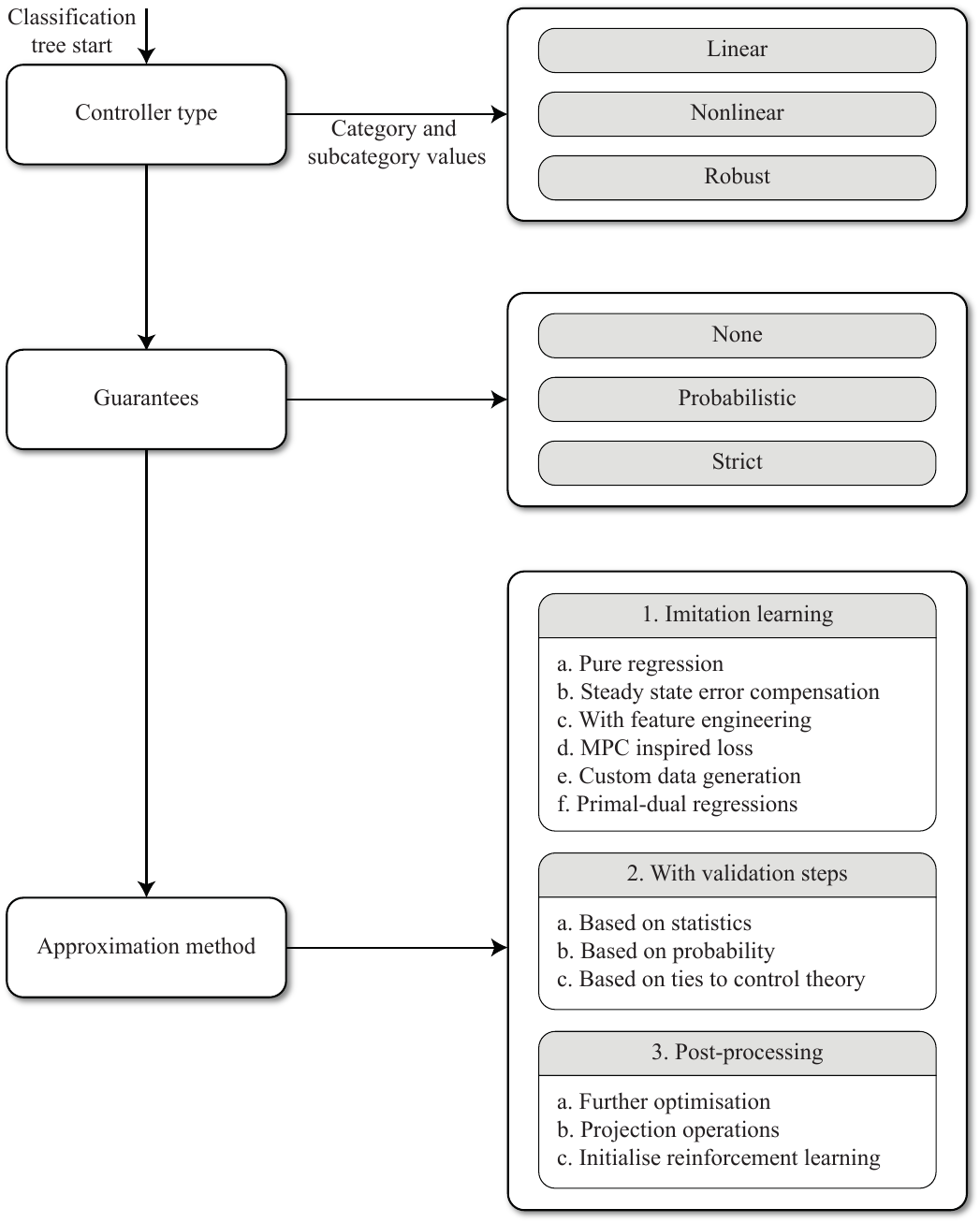}
    \caption{Proposed taxonomy of NNBO methods for MPC.}
    \label{fig:taxonomy}
\end{figure}

Numerous techniques have been put forth to incorporate NNBO into MPC controllers. This review classifies these methods according to the type of MPC they are suitable for, the guarantees they offer, and the approach employed to generate the approximation. Figure \ref{fig:taxonomy} summarises the proposed taxonomy. The controller type classification is straightforward, it indicates what type of MPC the framework is relevant to. However, the guarantees classification requires more explanation. In this context the term guarantees refers to the desirable theoretical properties of MPC controllers which, for instance, include constraint satisfaction, recursive feasibility, stability, and optimality. In an MPC controller such properties are guaranteed by design if the controller is adequately tuned and the optimisation solver uses hard constraints. However, when NNBO is used to approximate the solver, these guarantees are lost unless the approximation framework is designed to preserve them. Consequently, we differentiate between the cases where a framework is designed to preserve no guarantees, probabilistically preserve some or all guarantees, and strictly preserve some or all guarantees. The approximation method classification indicates the type of approach used to generate the neural approximation of the controller. For this field we provide values for three main categories and several subcategories thereof, all of which were identified from the works considered in this review. The alphanumeric labels of categories and subcategories shown in the figure are used for reference later in the text. Next, we elaborate on the meaning of method classifications to ease understanding of subsequent discussions.

\subsubsection{Approximation via imitation learning}

We refer to approaches where training data is fitted with an NN in the efforts of minimising a regression metric, such as MSE or RMSE, as imitation learning methods, provided that the resulting network is applied as is to replace the original controller after training. This provision is relevant because most methods start with, or use, imitation learning, but also apply other steps which later differentiate them. Our definition of imitation learning includes all types of network architecture, layer, activation function, or training algorithm. It is also broad enough to encompass the sub cases listed hereafter.  
\begin{itemize}
    \item \textbf{Pure regression}: When an NN is used to fit training data generated by sampling the original controller many times and the result is applied as is to replace the controller \citep{kumar2018deep, parisini1995receding, lucia2020deep,  lohr2019mimicking, lucia2018deep, bonzanini2021fast}.
    \item \textbf{Steady state error compensation}: When additional steps are taken to ensure that an approximation trained via pure regression does not suffer from steady state error or cause divergence from a steady state \citep{aakesson2005neural, chan2021deep, li2022using}.
    \item \textbf{With feature engineering}: When the training data generated by running the original controller is used to craft artificial features that serve as inputs to the NNBO approximation. The training still occurs via pure regression and the crafting of features can include the use of inputs that are not ever presented to the original MPC controller \citep{drgovna2018approximate, ortega1996mobile, vaupel2019artificial, nurbayeva2022deep, karg2018deep, lovelett2020some, parisini1998nonlinear}. 
    \item \textbf{MPC inspired loss}: When the regression error metric of the fitting problem is set to a function similar to the cost function of the original MPC controller \citep{aakesson2006neural, drgovna2022differentiable, ahmed1998neural}. 
    \item \textbf{Custom data generation}: When the training data for pure regression is generated through more intricate manners than iterative evaluation of the original controller at random points. For instance, via evaluation of the original controller over trajectories in closed-loop form, or by accounting for state estimation errors, etc \citep{cavagnari1999neural, karg2019learning}. 
    \item \textbf{Primal-dual regressions}: When the dual of the optimisation problem in the original controller is used or considered during NNBO training in an effort to exploit the properties of duality \citep{wang2022learning}.
\end{itemize}

\subsubsection{Approximation with validation steps}

These are approaches where trained approximations are put through tests to verify if specific performance requirements are met. Meeting said requirements must grant the approximation some form of guarantee, therefore, the standard practice of verifying the performance of a trained network on a validation or test set is excluded. From the works reviewed here, we have identified three sub categories that are relevant. Namely the cases where performance requirements are derived from results in statistics \citep{zhang2020near}, probability \citep{hertneck2018learning, nubert2020safe, karg2021probabilistic}, or ties to control theory \citep{karg2020efficient, karg2020stability, fabiani2022reliably, pin2010approximate, pin2013approximate}.

\subsubsection{Approximation with post-processing}

These are frameworks which coerce the outputs of approximations generated via imitation learning in an effort to preserve some guarantee or improve performance. The coercing action can take several forms such as using the output of the approximation to hot start subsequent optimisation \citep{chen2022large, lohr2020machine, vaupel2020accelerating, daosud2019efficient}, projecting the outputs onto sets that guarantee certain properties \citep{chen2018approximating, bonzanini2020toward, paulson2020approximate}, or using the approximation itself to kickstart the training of an RL policy \citep{karg2021reinforced}. We do not include the simple case of output saturation here because it is frequent practice in all other categories.

\section{Neural Network Based Optimisation for Linear Model Predictive Control}
\label{sec:NNBO4LMPC}

In this section we cover methods and frameworks for generating approximate LMPC controllers with NNBO. First, we discuss methods that provide no guarantees, then methods that provide probabilistic guarantees and finally methods that provide strict guarantees. Table \ref{tab:NNBO4LMPCSummary} provides a summary of all the methods that are covered alongside important metrics from case studies that can help the reader when choosing a method for their own application. In Table \ref{tab:NNBO4LMPCSummary} we characterise the size of an MPC problem based on the number of states $(n_x)$, outputs $(n_y)$ and inputs $(n_u)$ of the plant, and the length in steps of the prediction $(N_p)$ and control $(N_c)$ horizons. Where available we report speed up factors which indicate how much faster the resulting approximation is compared to the original controller as reported by the authors of the methods in case studies. We also report the validation environment where each method was tested, either through simulation (Sim) or on a real robotic system (Robot). Lastly, in the Label (L) column, we provide the classification under which each study has been categorised for the approximation method field of our proposed taxonomy.

\begin{table}[H]
\caption{Summary of methods and frameworks for generating approximate LMPC controllers that use NNBO.}
\label{tab:NNBO4LMPCSummary}
\resizebox{\linewidth}{!}{%
\begin{tabular}{@{}ccp{6cm}p{4cm}ccccccc@{}}
\toprule

Ref & L & \multicolumn{1}{c}{Core Concept} & \multicolumn{1}{c}{Guarantees} & $n_x$ & $n_y$ & $n_u$ & $N_p$ & $N_c$ & \multicolumn{1}{c}{Speed up} & Validation \\ 

\midrule
\cite{aakesson2005neural} & 1b & Train shallow NN to imitate LMPC using RMSE loss function, remove setpoint tracking error manually by zeroing NN outputs for setpoints. & None & 2 & 1 & 1 & 20 & 20 & 4,000 - 5,000x & Sim \\
\cite{drgovna2018approximate} & 1c & Train DNN with MSE loss to imitate MPC. DNN inputs defined via PCA and manual selection. & None & 286 & 6 & 6 & 22 & 22 & 7.6x & Sim \\
\cite{kumar2018deep} & 1a & Approximation of MPC with NN that combines LSTM and DNN. LSTM learns dependence on past actions while DNN learns dependence on current states. & None & 4 & 1 & 1 & & & & Sim \\
\cite{zhang2020near} & 2a & Fit DNN to primal and dual problems thereby enabling online constraint satisfaction check. If violations occur, call backup controller. & Probabilistic optimality with ability to check constraint satisfaction online. & 4 & 3 & 3 & 3 & 3 & 10x & Robot \\
\cite{chen2022large} & 3a & Train DNN to initialise active set solver and thereby reduce iterations to convergence. & Recursive feasibility and asymptotic stability. & 36 & 36 & 9 & 50 & 50 & 2x & Sim \\
\cite{chen2018approximating} & 3b & Project the outputs of an approximate DNN solver onto feasible output sets such that recursive feasibility is maintained. Train the DNN with policy gradient RL instead of supervised learning. & Recursive feasibility of inputs and states. & 4 & 4 & 2 & 10 & 10 &  & Sim \\
\cite{karg2020efficient} & 2c & Paper provides requirements for exact approximation of Explicit LMPC with DNN. & All EMPC guarantees for exact fits. Only constraint satisfaction for approximations. & 4 & 1 & 1 & 10 & 10 & & Sim \\
\cite{karg2020stability} & 2c & Verify input and state constraint satisfaction by solving a MILP that performs output range analysis on DNN ReLU approximation of solver, modify the approximation to behave like an LQR controller near the equilibrium point. & Input and state constraint satisfaction and stability. & 2 & 1 & 1 & 3 & 3 & & Sim \\

\end{tabular}%
}
\end{table}
\begin{table}[H]

\resizebox{\linewidth}{!}{%
\begin{tabular}{@{}ccp{6cm}p{4cm}ccccccc@{}}

Ref & L & \multicolumn{1}{c}{Core Concept} & \multicolumn{1}{c}{Guarantees} & $n_x$ & $n_y$ & $n_u$ & $N_p$ & $N_c$ & \multicolumn{1}{c}{Speed up} & Validation \\ 

\midrule
\cite{fabiani2022reliably} & 2c & Derives conditions based on worst-case error and Lipschitz constant under which DNN ReLU approximations are stable. Formulates MILPs to calculate such quantities. & Optimality, stability, and recursive feasibility. & 8 & 8 & 3 & 5 & 5 & & Sim \\
\bottomrule

\end{tabular}%
}
\end{table}

\subsection{Methods without guarantees}
\label{sec:NNBO4LMPC_none}

Herein we present and compare noteworthy methods to generate LMPC approximations that use NNBO but ignore the complexity of preserving the theoretical properties and guarantees of MPC. Methods of this kind have been successfully validated in simulation for small to very large plant models and can speed up MPC by several orders of magnitude. In the following discussion we introduce methods in order of ascending intricacy making it easy to see how the subcategories in our imitation learning label in Figure \ref{fig:taxonomy} arise. 

The simplest way of generating an NNBO approximation of MPC is to train a NN to mimic the outputs of a conventional optimiser via imitation learning. This is a type of regression problem where the inputs of the network include some or all of the inputs of the optimiser, such as current system states and reference trajectories. The regressed outputs are the optimal control actions in the first step of the control horizon calculated by the conventional optimiser. This idea was first proposed by \cite{parisini1995receding} for an unconstrained receding horizon regulator with a nonlinear plant model. Later on, \cite{aakesson2005neural} applied this methodology to LMPC. They used a shallow NN with hyperbolic tangent activation and trained the network using the Levenberg-Marquardt (LM) algorithm on data generated by solving the LMPC on closed loop trajectories corresponding to step changes in setpoints. One of the issues of this approach is that due to approximation error, the outputs of the resulting NN can be non-zero at setpoints and when this occurs, the system starts to diverge from the setpoint. \cite{aakesson2005neural} addressed this issue by manually zeroing the network outputs when all state derivatives are zero and the current output is equal to the desired setpoint. The effectiveness of the approach was demonstrated via simulation and the resulting approximate LMPC was 4,000 -- 5,000 times faster than the original LMPC controller. \cite{drgovna2018approximate} showed that this approach could be extended to large problems with hundreds of state variables. In their work a Deep NN (DNN) was used instead of a shallow network and the inputs to the network were selected using expert knowledge of the system, Principal Component Analysis (PCA) and feature crafting. The study showed that the input features of the network can extend beyond what is available to the conventional optimiser in order to ease or increase the accuracy of the DNN regression. Similarly, \cite{kumar2018deep} showed that more advanced network architectures could be used to achieve the same purpose. Their framework combines a feedforward DNN with a network with Long Short-Term Memory (LSTM) units. The feedforward network learns the dependence of outputs on current states and the LSTM network learns the dependence of outputs on previous outputs similar to how a conventional solver is started from previous solutions. The networks are trained together and their output is fused with a weighted average. The authors showed through case studies that the proposed approach was a better regression than either network would produce independently. However, only a very small problem was considered, as shown in Table \ref{tab:NNBO4LMPCSummary}, and no details of computation time improvements with respect to the original controller were provided. Therefore, it is not clear if this method generalises well to larger problems or if it enables real time MPC, but the idea of using more complex neural units and network architectures is noteworthy. 

The common problem of approaches without guarantees is that safety of the controller cannot be ensured by design. Despite their differences, all aforementioned approaches boil down to solving a regression problem and although NNs are universal function approximators \citep{barron1993universal}, exact regressions are impractical or intractable in practice and therefore regression error is unavoidable. With a sufficiently complex network and enough training data, it is possible to minimize the regression error substantially. However, as highlighted by \cite{aakesson2005neural}, this error can grow over time when the network is engaged in closed-loop control. Ultimately, the performance of the approximate LMPC controller may deviate significantly from that of the exact controller. Hence, approaches without guarantees require the use of add-on safety mechanism that run in parallel to the controller and perform functions such as control action saturation, constraint violation detection, instability detection, and emergency stopping. This arrangement is acceptable for some practical applications such as controlling inexpensive unmanned systems operating in low risk environments. However, for more critical systems the theoretical guarantees of MPC are desirable.

\subsection{Methods with probabilistic guarantees}
\label{sec:NNBO4LMPC_probabilistic}

When a conventional optimisation solver converges to a solution, that solution is guaranteed to comply with all the constraints of the problem and be locally or globally optimal, within the tolerance limits stipulated by the settings of the solver. \cite{zhang2019safe, zhang2020near} proposed a means to replicate this behaviour with high probability by training DNNs to imitate the output of conventional solvers for both the primal optimal control problem and its corresponding dual. Based on the properties of duality, the resulting primal and dual approximations can be used to conduct suboptimality and constraint satisfaction checks at runtime. If requirements are not met, a backup controller can be used instead of the calculated control actions. The study showed how the training problems can be formulated as a chance-constrained optimisation problem which can be approximated through randomisation. Thus, if the number of training samples is chosen adequately, both the primal and dual approximations can be trained to meet constraints and prescribed suboptimality thresholds with high probability. Statistical guarantees on the approximate randomised problem are purely dependent on the number of random samples used to solve the problem and the number of samples can be computed from scenario optimisation theory for convex problems, and from statistical learning theory for non-convex problems. The process of training a DNN approximation falls under the latter category (i.e. non-convex approximate chance-constrained problems). Since the calculated number of samples required to provide guarantees is generally very large, a heuristic can be adopted where a subset of samples is used for training and then the trained DNN is tested on all remaining samples. By following this framework one can train fast DNN based approximations of LMPC which comply with the input, output, and state constraints of the exact optimal control problem with high probability, and which yield results that meet user prescribed suboptimality bounds with high probability. Most importantly, the approach also enables the detection of instances where probabilistic guarantees have failed during operation. This can be used to trigger the activation of a backup control strategy, thereby ensuring the continuity of safe operation. The framework was demonstrated on an LVMPC integrated chassis control problem and the study reported a 10x speedup on the target hardware used in the automotive application. These results are very positive but it should be noted that the length of the horizon considered in the study was only three time steps and the underlying plant only had four states and three outputs. Yet, despite the small dimensions of the problem, the primal and dual DNN approximations had output dimensions 9 and 36, respectively. For larger plants and controllers with longer horizons, the ratio of inputs to outputs of the DNNs would be very low making the regression problem increasingly difficult and likely infeasible. Therefore, scalability of this method to larger problems is unlikely. 

There are no other noteworthy methods that can be used to produce LMPC approximations with probabilistic guarantees but similar approaches have been developed for other types of MPC as will be discussed in Sections \ref{sec:NNBO4NMPC_probabilistic} and \ref{sec:NNBO4RMPC_probabilistic}. Methods discussed in those sections have been shown to scale to larger problems and in principle, such methods can also be applied to LMPC. In the next section we turn our attention to frameworks that provide strict guarantees. 

\subsection{Methods with strict guarantees}
\label{sec:NNBO4LMPC_strict}

Safety guarantees that hold with high probability are better than no guarantees at all, but ideally one would want an MPC approximation that is a drop-in replacement for the original controller but runs faster and preserves all of its desirable features, in other words, one would want an approximation with strict guarantees. Enabling this idealised scenario is difficult because there are no direct ties between MPC theory and NNs. In this section we present frameworks that circumvent this problem by using strategies like hot starting conventional solvers from approximations and using projection operators to coerce approximations. We also cover several works on the development of ties between LMPC theory and NN approximations with specific architectures.

\cite{chen2022large} proposed an explicit-implicit MPC approach where a DNN optimiser is used to warm start an active set solver. With this setup all MPC guarantees are preserved because a conventional solver is still used at every control iteration. The DNN optimiser is based on a primal formulation where the output of the network contains the primal variables for the entire prediction horizon and the loss function is based on the error between the values of the Lagrangian for the predicted and real solutions. By themselves the outputs of the network have no guarantees but because all primal variables are predicted, a simple algebraic check can be performed to assess the feasibility and suboptimality of the solution. If the solution is infeasible or suboptimal beyond some acceptable threshold then the output of the DNN is used to warm start an active set solver that coerces the solution to feasible and optimal values. Through simulation studies involving large plants and long horizons, \cite{chen2022large} showed that initialising the active set method with the solution found by the DNN significantly sped up the computation compared to the case where the active set method was cold started. However, cold starts are normally only required on the first iteration of MPC, on subsequent iterations warm starts can be performed using the previous solution. Therefore, this approach only benefits controllers that can already run in real time. Later on, \cite{chen2018approximating} proposed a different approach in which the outputs of a DNN approximation are projected onto feasible sets using Dykstra's projection. This coercion preserves the recursive feasibility guarantees of MPC but requires significant overhead to perform the projection, and computation of the maximal control invariant set which is only feasible for some problems. For this study the authors demonstrated the method via simulation for a small plant and did not report on computation time improvements. 

Regarding the development of theoretical ties between MPC and NN approximations thereof, Karg and Lucia presented seminal results for approximations of EMPC \citep{karg2020efficient} and LMPC \citep{karg2020stability}. In \cite{karg2020efficient}, the authors derived the requirements that a DNN with Rectified Linear Unit (ReLU) activation functions needs to satisfy to exactly fit an EMPC policy. They also suggested methods for quantifying the approximation error of networks that did not meet these requirements. Then, in \cite{karg2020stability}, the authors started developing similar theory for LMPC. Specifically, they designed a MILP that can be applied to each output of a DNN ReLU approximation of LMPC to perform output range analysis and determine if the output in question can violate its constraints during closed loop operation. In practice generating an approximation that passes the MILP checks for all outputs can be difficult and therefore the authors also proposed a method to modify trained approximations via optimisation and make them behave like an LQR controller near the equilibrium region. In both studies, \cite{karg2020efficient, karg2020stability} proved the effectiveness of their methods via simulation with small plants with a single input and a controller with a short horizon as shown in Table \ref{tab:NNBO4LMPCSummary}. For LMPC controllers of that size, conventional optimisers can run in real time at thousands of $Hz$ and because the methods proposed by Karg and Lucia have not been demonstrated on larger problems their scalability is unclear. Most recently, \cite{fabiani2022reliably} built on the foundations by \cite{karg2020efficient} and proposed an algorithm involving two MILPs to fit a minimal complexity DNN ReLU that approximates a given LMPC controller and preserves strict stability and constraint violation guarantees. The output of the first MILP is the upper bound on approximation error of the DNN in terms of the difference in computed optimal inputs with respect to a real optimiser. The solution of the second MILP is the Lipschitz constant associated to the approximation error. If the constant is below a certain threshold then the DNN outputs are guaranteed to ensure closed-loop stability around the origin of the plant. Importantly, this framework always requires solving two MILPs while the method by \cite{karg2020stability} requires solving a MILP per network output. \cite{fabiani2022reliably} also noted that while MI optimisation typically has poor scalability with problem size, their numerical experiments showed weak dependence of MILP solution times on number of plant states and strong dependence on number of inputs. However, their experiments used a prediction horizon of only five steps and the scalability of the method with prediction horizon length was not discussed. Overall, good progress has been made in developing NN based approximations of LMPC with strict guarantees but such methods have only been shown to work in small problems and their scalability has not been explored in depth.

\section{Neural Network Based Optimisation for Nonlinear Model Predictive Control}
\label{sec:NNBO4NMPC}

In this section we cover methods and frameworks for generating approximate NMPC controllers with NNBO. As before, we discuss methods that provide no guarantees, probabilistic guarantees and strict guarantees separately and provide a summary of all methods in Table \ref{tab:NNBO4NMPCSummary} along with performance metrics on case studies.

\begin{table}[H]
\caption{Summary of methods and frameworks for generating approximate NMPC controllers that use NNBO.}
\label{tab:NNBO4NMPCSummary}
\resizebox{\linewidth}{!}{%
\begin{tabular}{@{}ccp{6cm}p{4cm}ccccccc@{}}
\toprule

Ref & L & \multicolumn{1}{c}{Core Concept} & \multicolumn{1}{c}{Guarantees} & $n_x$ & $n_y$ & $n_u$ & $N_p$ & $N_c$ & \multicolumn{1}{c}{Speed up} & Validation \\ 

\midrule

\cite{parisini1995receding} & 1a & First paper to propose the idea of training an NN to imitate MPC via pure regression. & None & 6 & 3 & 2 & 30 & 30 & & Sim \\
\cite{ortega1996mobile} & 1c & Train a shallow NN to imitate MPC using RMSE loss. Inputs to the network include past inputs and sensor readings that are not fed to MPC. & None & 5 & 3 & 2 & 7 & 6 & & Robot \\
\cite{cavagnari1999neural} & 1e & Train shallow NN with sigmoid activation to imitate LMPC, generate training data from LQR solution as well as MPC. & None & 4 & 4 & 1 & 10 & 10 & & Robot \\
\cite{lucia2020deep} & 1a & Train $tanh$ DNN with MSE loss to imitate MPC, then implement the approximate controller using FPGA for much faster execution. & None & 2 & 2 & 2 & 10 & 10 & 500,000x & HIL \\
\cite{chan2021deep} & 1b & Train DNN with MSE loss to imitate MPC, then add steady state error correction term to loss function and retrain to generate offset free DNN. & None & 2 & 1 & 1 & & & & Sim \\
\cite{vaupel2019artificial} & 1c & Train DNN to imitate MPC using MSE loss, use measurable disturbance variables as network inputs. & None & 7 & 7 & 1 & 10 & 8 & 10x & Sim \\

\end{tabular}%
}
\end{table}
\begin{table}[H]
\resizebox{\linewidth}{!}{%
\begin{tabular}{@{}ccp{6cm}p{4cm}ccccccc@{}}

Ref & L & \multicolumn{1}{c}{Core Concept} & \multicolumn{1}{c}{Guarantees} & $n_x$ & $n_y$ & $n_u$ & $N_p$ & $N_c$ & \multicolumn{1}{c}{Speed up} & Validation \\ 

\midrule
\cite{nurbayeva2022deep} & 1c & MPC motion planner approximated by DNN via pure regression, explores input dimensionality reduction with autoencoders. & None & 6 & 6 & 6 & 10 & 10 & & Robot \\
\cite{lohr2019mimicking} & 1a & Train DNN to imitate MPC with binary control inputs using RMSE loss function. & None & 3 & 3 & 3 & 24 & 24 & 12.12x & Sim \\
\cite{karg2018deep} & 1c & Train DNN to imitate MPC with binary control inputs. Include trajectories of disturbances over horizon as part of the training data. & None & 4 & 4 & 4 & 24 & 24 & & Sim \\
\cite{lovelett2020some} & 1c & Use manifold learning to map augmented state space to optimal control inputs. Then train DNN to retrieve optimal control inputs from manifold space. & None & 2 & 1 & 1 & 20 & 20 & 36,000,000x & Sim \\
\cite{aakesson2006neural} & 1d & Train a shallow $tanh$ NN with an MPC loss function on data corresponding to closed loop trajectories. Use result to replace conventional optimiser. & None & 4 & 2 & 2 & 60 & 25 & & Sim \\
\cite{drgovna2022differentiable} & 1d & Create and train DNN plant based on state space architecture. Train DNN controller with loss function inspired on MPC cost. & None & & 1 & 1 & 32 & 32 &  & Robot \\
\cite{li2022using} & 2b & Formulate training data generation and DNN approximation training as a single optimisation problem. & Probabilistic constraint satisfaction. & 4 & 2 & 2 & 20 & 20 & 1,000x & Sim \\
\cite{parisini1998nonlinear} & 2c & First formalised framework for generating shallow NN approximations of NMPC. Authors propose methods to compute NN size and terminal NMPC costs that make NN feasible. & Bound output error and stability to region near origin. & 2 & 2 & 1 & 100 & 100 & & Sim \\
\cite{pin2010approximate} & 2c & Further investigation of approximation properties of the framework proposed in \citep{parisini1998nonlinear}. & Practical stability and input constraint satisfaction. & 2 & 2 & 1 & 8 & 8 & & Sim \\
\cite{pin2013approximate} & 2c & Extend \citep{pin2010approximate} with requirements that approximations have to meet to guarantee stability and constraint satisfaction. & Practical stability and constraint satisfaction on inputs and states. & 2 & 2 & 1 & 8 & 8 & & Sim \\
\end{tabular}%
}
\end{table}
\begin{table}[H]
\resizebox{\linewidth}{!}{%
\begin{tabular}{@{}ccp{6cm}p{4cm}ccccccc@{}}

Ref & L & \multicolumn{1}{c}{Core Concept} & \multicolumn{1}{c}{Guarantees} & $n_x$ & $n_y$ & $n_u$ & $N_p$ & $N_c$ & \multicolumn{1}{c}{Speed up} & Validation \\ 

\midrule
\cite{ahmed1998neural} & 1d & Create a first order differentiable, MPC-inspired loss function and use it train a DNN that imitates MPC. Derivatives of loss are computed using block partial derivatives. & Input constraint satisfaction via saturation and stability at equilibrium point. & 2 & 1 & 1 & 300 & 300 & & Sim \\
\cite{lohr2020machine} & 3a & Train DNN to predict first value of a binary control input and thereby enable conversion of MIQP into QP. & Constraint satisfaction and input feasibility via solution of QP at every iteration. & 3 & 3 & 5 & 48 & 48 & 3x & Sim \\
\cite{vaupel2020accelerating} & 3a & Train DNN with MSE loss to imitate MPC then use it to hot start full NMPC solution or alongside a post processing QP that corrects the outputs of the network and makes them constraint compliant. & All MPC guarantees when DNN is used to hot start solver. Good practical constraint satisfaction when post processing QP is used. & 6 & 5 & 3 & 7 & 5 & 24x & Sim \\
\cite{bonzanini2020toward} & 3b & Train a DNN to imitate NMPC, project outputs of DNN onto RAI which in turn will ensure that states stay within RCI. The projection operation can be precomputed as the solution of an mpQP. & Input and state constraint satisfaction. & 2 & 2 & 2 & 100 & 100 & 25 - 90x & Robot \\
\cite{wang2022learning} & 1f & Integrate Lagrange duality into DNN loss function to guarantee constraint satisfaction of the primal problem, train the DNN approximation through imitation learning on the modified loss function. & Input and state constraint satisfaction. & 3 & 3 & 2 & 50 & 50 & 6 - 72x & Sim \\
\bottomrule

\end{tabular}%
}
\end{table}

\subsection{Methods without guarantees}
\label{sec:NNBO4NMPC_none}

As for LMPC, the approximation methods without guarantees for NMPC are predominantly of the imitation learning kind. The exact way in which the training is implemented can take many forms because there are many possible inputs, outputs, network architectures and loss functions that one may choose to use. We hereby present a selection of unique frameworks and methods which cover all avenues that have been explored in literature so far. 

\cite{parisini1995receding} were the first to propose the idea of training an NN to imitate the outputs of an NMPC controller. Their framework consisted of fitting a shallow NN with $tanh$ activation to training data generated by repeatedly applying a stabilizing NMPC controller to initial states sampled randomly from a bound set. The inputs to the network were the current system states and the outputs were the control actions calculated by the NMPC controller for the first step in the control horizon, an MSE loss function and a gradient algorithm were used to solve the regression problem. The effectiveness of the approach was demonstrated via simulation on two example problems one of which included a MIMO plant of medium size as reported in Table \ref{tab:NNBO4NMPCSummary}. \cite{ortega1996mobile} extended these results by applying the same method to a path tracking NMPC controller used as the navigation algorithm of a mobile robot. In this case the shallow NN was provided with additional inputs such as a parametric representation of the reference trajectory, distance sensor readings for obstacle detection, and most notably, the control actions at the previous time step. \cite{ortega1996mobile} validated their approximate controller with experiments on the real robot, thus showing, for the first time, that NNBO can indeed facilitate the deployment of NMPC controllers that would otherwise be unable to execute in real time. \cite{cavagnari1999neural} further proved this point with their experimental results on a seesaw mechanism. Using the same random sampling and MSE loss framework, they replaced a stabilizing NMPC controller with an NN approximation capable of running in real time. Moreover, they introduced a small novelty by training the NN with data generated from two different controllers operating in different regimes. For samples near the equilibrium position, they used training data generated from an LQR controller with a linearised plant, for points outside this region, training data was generated using the NMPC controller. The result was an approximate NN controller that could seamlessly transition between two control behaviours for better stability. The three aforementioned studies were conducted in the late 90s. However, following their publication, the simple imitation learning approach received little attention for almost two decades. 

Interest in the simple imitation learning approach reignited around 2018, with many researchers proposing modifications and extensions of the method. \cite{lucia2020deep} showed that the $tanh$ DNN approximation of an NMPC controller could be implemented in Field Programmable Gate Array (FPGA) hardware to achieve much faster execution of the approximate controller and thus enable the control of systems with ultra fast dynamics. The effectiveness of this approach was demonstrated on a resonant power converter application including Hardware In the Loop (HIL) testing. From these tests, \cite{lucia2020deep} reported a speed up factor of approximately half a million times for their most optimised FPGA implementation compared to the original NMPC controller. 

\cite{chan2021deep} proposed an extension of the basic framework to eliminate steady state setpoint tracking error due to approximation error. In their framework, a DNN approximation of an NMPC controller is first generated by solving the usual regression problem with an MSE loss function. Then, the approximation and original controller are applied to setpoint tracking trajectories to generate a new data set. This dataset is used to retrain the network using a modified loss function that includes an approximation error term. Retraining on this modified loss function effectively minimises the approximation error which in turn minimises, or effectively removes, setpoint tracking error. \cite{chan2021deep} also implemented their approximate controller on FPGA hardware and demonstrated the effectiveness of their approach via simulation for a small problem. Alternatively, \cite{vaupel2019artificial} showed that acceptable setpoint tracking performance can also be achieved by including disturbance measurements in the inputs of the neural network approximation and training with the pure regression setup on an MSE loss function. For applications where disturbances can be measured, this approach is significantly simpler compared to the method proposed by \cite{chan2021deep}. 

For complex systems with difficult to parameterise inputs, \cite{nurbayeva2022deep} proposed the use of autoencoders for input dimensionality reduction. In their work, an NMPC motion planner for human-robot interaction tasks was replaced with a DNN approximation and successfully tested in simulation and on the real robot. Autoencoders were used to parameterise human motion inputs to the approximate controller, robot state inputs were fed directly. While the approach may be worth considering for some applications, the authors themselves point out that the resulting controller is not generalisable to motion scenarios beyond the specific interaction for which it was trained. The controller is also unable to cope with significant deviations from the interaction it was trained for.

\cite{lohr2019mimicking} and \cite{karg2018deep} showed that the basic imitation learning setup is also applicable to NMPC controllers with binary variables where the underlying optimisation problem is of the mixed-integer type. \cite{lohr2019mimicking} trained their approximation using an RMSE-based loss and the inputs to their network were the current states, references and disturbances of the system. \cite{karg2018deep} considered the same inputs but used the entire predicted disturbance trajectory over the prediction horizon instead of only the latest measurement. In both studies the effectiveness of the approach was demonstrated via simulation on heating system control problems with MIMO plants of similar size. The independence and success of these studies suggests that NNBO is a viable method for accelerating NMPC problems with binary control variables, but there is room for in-robot validation studies as well as studies that explore the more general problem of handling integer variables instead of only binary variables. 

In a significantly different approach, \cite{lovelett2020some} noted that the mapping function between states and optimal control actions (e.g. the function that the NN in previous approaches is trying to learn) can be complicated and difficult to learn because the augmented state space, which consists of plant states and controller references, is often a poor parametrisation of the control law. Motivated by this observation, they proposed the use of diffusion maps, a type of manifold learning algorithm, to learn a potentially lower dimensional parametrisation of the mapping function. If such a parametrisation exists, it is a simpler map between manifold space and control actions. However, using this map to obtain control actions given the augmented state is not straightforward. The authors addressed this challenge with two NNs. The first NN was used to learn the mapping between augmented state space and manifold space, and the second network learnt the mapping between manifold space and optimal control actions. In theory these regression problems are easier than directly approximating the original controller. The authors demonstrated the effectiveness of the approach via a simulation case study on a nonlinear system with a small plant. Reports on execution time improvements indicated that the neural approximation was 36,000,000 times faster than the original NMPC.

So far, we have discussed methods that make use of a simple imitation learning setup where error information is used to solve a regression problem via gradient descent, hereafter we present other alternatives. \cite{aakesson2006neural} proposed an approach wherein the error-based loss function is replaced with an MPC-like loss function. This new loss function encourages the network to learn the dynamics of the controller instead of over-fitting it for a bound set of inputs. In their study the authors formulated this method for trajectory tracking NMPC. The network inputs are those typically available to an NPMC controller such as current estimated states, measured outputs, reference outputs, and previous control actions. The network outputs are the control command increments to be applied at the current control interval. In this case training consists of minimising the MPC-like loss function over a large set of trajectories on which the network is applied in a closed loop setup. The minimisation task can be formulated as a nonlinear least squares problem and solved using a gradient algorithm. The main advantage of this method is that the size of the resulting optimisation problem is independent of the horizon length used in the MPC loss function. This means that the NN approximation can be trained with longer horizons, surpassing what would be practically feasible for deployment in the exact NMPC controller. 

\cite{drgovna2022differentiable} built further on the idea of departing from error-based loss functions in their Differentiable Predictive Control (DPC) framework where both the optimiser and plant model of the original controller are replaced with DNNs with architectures and loss functions inspired by MPC. More specifically, the plant model is approximated using a neural state space architecture which can learn the dynamics of nonlinear systems from recorded data. The main advantage of this approach is that the model directly learns the disturbed and uncertain dynamics of the system without additional effort. The trained plant can then be used to train a policy network using an MPC-inspired loss function and a dataset that emulates closed loop control. Crucially, the policy and plant networks can be combined into a single model that is continuously differentiable making it easy to train with any modern training algorithm such as ADAM \cite{kingma2014adam}. Again, the advantage of combining the policy and data-derived plant networks into a single model is that the resulting policy inherently compensates for disturbances and uncertainties, thereby making it comparable to RMPC. In their study \cite{drgovna2022differentiable} demonstrated the capabilities of DPC by applying it to a real SISO system with an unknown nonlinear plant. Experimental results showed that the performance of the controller was on par with that of an exact EMPC alternative. However, the DPC controller could handle much more complex tasks such as tracking dynamic references with dynamic constraints. Such results are promising and indeed demonstrate the ability of DPC to control nonlinear systems with disturbed, uncertain and unknown dynamics, however, the performance and scalability of DPC to MIMO systems are yet to be established.

\subsection{Methods with probabilistic guarantees}
\label{sec:NNBO4NMPC_probabilistic}

In this section, we discuss a method for generating NNBO approximations of NMPC with probabilistic guarantees on constraint satisfaction based on Hoeffding's Inequality (HI). We start by providing a brief explanation of this inequality because other studies discussed hereafter are also based on it. HI, as shown in \eqref{eq:hoeffdings}, defines the upper bound of the probability of the mean of a set of $n$ randomly sampled variables $(\xi_i)$ from a distribution being further than a value $(\epsilon)$ from the mean of the distribution itself $(E(\xi))$ \citep{von2011statistical}. The value of the upper bound is given by the RHS of \eqref{eq:hoeffdings}. Thus note that for a fixed $\epsilon$ the upper bound decreases with the number of samples. This statistical result can be applied in several ways when generating an NNBO approximation of MPC to yield different kinds of guarantees. For instance, it is possible to train DNN optimisers that find solutions with bounded levels of suboptimality with a certain probability, or DNN optimisers that guarantee constraint satisfaction with some probability, or DNN optimisers that guarantee stability.
\begin{equation}
    \label{eq:hoeffdings}
    P\left( \left| \frac{1}{n} \sum_{i=1}^{n}\xi_i - E(\xi) \right| \geq \epsilon \right) \le 2e^{-2n\epsilon^2}
\end{equation}
\cite{cao2020deep} and \cite{li2022using} used HI to come up with a framework for generating NNBO approximations of NMPC with probabilistic guarantees on constraint satisfaction. Similar to methods with MPC-inspired loss functions, their method embeds the MPC optimisation problem into the DNN training problem and in this case it results in a unified nonlinear two-stage stochastic optimisation problem. To provide the statistical guarantees, \cite{cao2020deep} used the same statistical learning theory principles as \cite{zhang2020near}, but instead of applying it to point samples they applied it to closed-loop trajectory samples. That idea was first proposed in the work by \cite{hertneck2018learning} which will be discussed in Section \ref{sec:NNBO4RMPC_probabilistic}. In this case an appropriate choice of number of samples leads to a probabilistic guarantee of constraint satisfaction for the DNN approximation for any trajectory starting from any point in the prescribed distribution of starting points. Similar to \cite{zhang2020near}, the calculated number of required samples is usually quite large. Therefore, \cite{li2022using} employed a heuristic based on HI where DNN approximations for an increasing number of sample trajectories are produced until satisfactory performance and guarantees are achieved. In this study two methods were proposed to solve the ``optimise and train" nonlinear two-stage stochastic optimisation problem, namely, a numeric interior point based solver, and an equivalent RNN training problem. Both methods were demonstrated on a water tank level control problem with nonlinear dynamics, four states, two outputs, and two inputs. The resulting DNN approximation sped up optimisation by three orders of magnitude compared to a conventional solver. \cite{li2022using} also showed how the performance of the approximation improves with the incorporation of additional trajectories and carefully selected critical scenarios during the training process. Importantly, as more samples are added, the interior point solver eventually becomes unable to solve the problem. As a result, for large data sets only the RNN optimisation approach is viable. The sole shortcoming of the study is not considering more than one benchmark problem to evaluate the scalability and robustness of the approach. 

\subsection{Methods with strict guarantees}
\label{sec:NNBO4NMPC_strict}

In this section, we cover methods and frameworks that can be used to generate NNBO approximations of NMPC in ways that strictly preserve some or all of the desirable properties of the original controller such as constraint satisfaction, stability and recursive feasibility. Following their pioneering work \citep{parisini1995receding} on learning MPC policies with NNs, Parisini et al. continued to develop their ideas and eventually addressed the problem of constrained NMPC. In \cite{parisini1998nonlinear}, they proposed a set of optimisation routines and heuristic iterative procedures that could be used to compute bounds on the terminal weights and number of neurons required to create NMPC approximations with bounded output error over the set of states for which the approximation was generated. With sufficiently low bounds on the output error the approximate NMPC controller is guaranteed to stabilise the system to a small user-prescribed region around the origin, thus making this approach the first with guarantees on stability. The proposed methods were demonstrated on a small plant. However, on a second example involving a larger and more complex plant, the authors omitted all of the procedures that yield the stability guarantees and went back to the pure regression approach, presumably because the problems involved were intractable for large systems. This line of work was continued over a decade later by \cite{pin2010approximate, pin2013approximate}, where the stability properties of the approximate control laws were further studied in the presence of approximation error and modelling uncertainties. As reported by \cite{parisini1998nonlinear}, these studies showed that the approximate control policies generated with this framework are only able to stabilise systems to a region near the origin. Furthermore, no new evidence on the scalability of the approaches was provided. Instead, both studies used the same example system, a nonlinear oscillator, featuring a small plant with two states and one input. The prediction and control horizons considered in the exact NMPC controllers were also short, as outlined in Table 2. In fact, \cite{hertneck2018learning} reported that for the framework in \cite{pin2013approximate} the admissible approximation error required to yield guarantees was typically unachievable.

In the same year when \cite{parisini1998nonlinear} formalised their framework, \cite{ahmed1998neural} proposed a fundamentally different approach, where instead of learning the NMPC policy via regression, the loss function of the approximating NN was set to the cost function of a regulating NMPC controller. The resulting NN could be trained using back propagation by calculating the required derivatives using the concept of block partial derivatives. The authors also introduced an important constraint to ensure stability at the origin. Specifically, they set out to train a network that would produce no output if the system was at the equilibrium position. As they point out, a linear control law with no offset achieves this by design, so the architecture of the approximating NN was set to a deep NN with no bias terms and $tanh$ activation such that if all inputs to the network were zero, the outputs would also be zero. The proposed approach was shown to work for state and output regulation tasks of small linear and nonlinear plants with prediction and control horizons of up to 300 steps, with and without input disturbance. While those simulation results are very promising, it should be noted that the approach has some disadvantages. For instance, training requires human supervision at every step to ensure stability. This involves manually adjusting the network architecture and learning rate as well as gradually increasing the length of the prediction horizon. Also, the only guarantees that the approach by \cite{ahmed1998neural} provides are input constraint satisfaction via saturation and stability at the equilibrium point. State and output constraint satisfaction are not guaranteed and neither is stability during regulation. 

Some works have focused on the approximation of NMPC controllers with binary control inputs. \cite{lohr2020machine} considered the problem of mixed-integer MPC with binary decision variables for linear state space plant models. Despite considering a linear state space plant, the use of binary control variables makes this problem nonlinear. Typically, this type of MPC would require solving a Mixed Integer Quadratic Program (MIQP) which is much slower than solving a standard QP. To get around this, the authors proposed training a NN to predict the value of all binary variables at the first step in the prediction horizon. Moreover, they relaxed the binary constraint for all other steps in the prediction horizon by allowing the binary variables to be continuous but limited to the range $[0,\;1]$. This effectively converted the MIQP into a standard QP that could be solved much quicker. Given that the resulting approximate strategy still involves solving a QP in real time, this approximation framework preserves all the theoretical properties of LMPC. \cite{lohr2020machine} demonstrated the effectiveness of the approach on a domestic heating control problem with a small plant and two binary control inputs. For that problem, the performance of the approximate controller was on par with that of the original controller where the MIQP was solved using a conventional solver. Nonetheless, the approximate controller ran three times faster. 

\cite{vaupel2020accelerating} further explored the concept of using NN approximations alongside conventional optimisation methods. They studied two strategies: initialising the NMPC solution from the outputs obtained from a DNN approximation of the controller, and refining the solution of DNN approximations by solving a QP that approximates the original NMPC problem. Both of these approaches require training of a DNN approximation of the original NMPC controller. In this case the approximation needs to predict the value of all control actions at each step along the control horizon of the controller. To achieve this the authors used an imitation learning setup with an MSE loss function and training data generated from running closed loop simulations of the NMPC controller. The inputs of the network were set to the states, references, and disturbances at the current control interval. In the NMPC initialisation case, the control actions produced by the DNN approximation were used to hot start a conventional optimisation solver which went on to solve the NMPC problem to convergence. In the second case, the outputs of the DNN were used to construct a QP that was an approximation of the full NLP problem. This approximation was generated by computing the hessian of the Lagrangian and gradients of the objective and constraint functions for the control action values predicted by the DNN approximation. The QP was designed such that its solution satisfied a first order approximation of the constraints of the problem thereby strictly preserving the theoretical guarantees on MPC. In practice guaranteeing the convexity and feasibility of the resulting QP requires additional strategies such as modifying the calculated Hessian until it is positive definite or solving a feasibility restoring QP. In both cases the output of the post processing QP is a set of deltas that can be added to the original guess of the DNN approximation to improve it and make it constraint compliant. Through simulation studies on a small MIMO plant and an original NMPC controller with short prediction and control horizons, \cite{vaupel2020accelerating} reported that using the DNN outputs to initialise the NMPC solution slowed down the controller by about 45\% on average. However, the maximum solution time decreased by 42\% and computation times were much more consistent. Thus, while this approach could be useful for some applications, it is not relevant for the purpose of significantly speeding up existing controllers. On the other hand, the post processing QP approach was on average 24 times faster than solving the full NMPC problem and it led to negligible to no deterioration in control performance. When compared against the popular approach of generating an NMPC approximation via pure regression, the resulting strategy was able to respect all control constraints and maintain close to optimal cost values during operation. In contrast, the simple regression approach violated constraints and exhibited running costs far from optimal.

\cite{bonzanini2020toward} proposed another post processing method relying on projection operations instead of further optimisation. In this method the outputs of a DNN approximation of NMPC are projected onto the Robust Admissible Input (RAI) set. This projection ensures adherence to all input constraints and guarantees that system states remain within the Robust Control Invariant (RCI) set, which is state constraint compliant by design. The projection operation consists of solving a secondary optimisation problem at every control interval, but said optimisation problem can be formulated as a standard QP and solved offline as a multi-parametric QP (mpQP). The problem with this approach, as with other of similar nature, is that mpQP does not scale well to large problems and for large cases the lookup operation itself can be infeasible in real time at high control rates, especially for embedded systems where computational resources are limited. Indeed, the example provided by \cite{bonzanini2020toward} was for a small plant with only 2 states and 2 inputs, but it should be noted that the length of the prediction and control horizons was quite significant at 100 steps. Thus, while the scalability of the method with plant size is unknown, for small plants the method scales well to long horizons.

Instead of post-processing the output of a DNN approximation, \cite{wang2022learning} proposed an imitation learning setup where the resulting network respects the constraints of the original NMPC problem by design. This result was achieved by setting the loss function of the DNN to a function corresponding to the dual of a problem which minimises the difference in cost between the approximate and original controllers subject to the constraints of the original controller. The training problem then becomes a max-min problem where the goal is to maximise the dual variables while minimising the loss by adjusting the parameters of the DNN. Wang et al. solved this problem using dual gradient descent and a heuristic algorithm that progressively expanded the size of the training data set until a desired level of performance was achieved which guaranteed input and state constraint compliance of the approximation. 

The review of all aforementioned methods shows that strictly preserving the guarantees on NMPC is a much more difficult task than it is for LMPC. There are options to preserve input and state constraint satisfaction, or options to guarantee stability to a region near the origin, or options to preserve stability at equilibrium points, but besides the approaches involving further optimisation, there are no frameworks that can unconditionally preserve all desirable guarantees at once. The development of such a framework remains an open research question of high importance, especially given that a purely NN based approach could significantly outperform further optimisation approaches in terms of computation time reductions and scalability.

\section{Neural Network Based Optimisation for Robust Model Predictive Control}
\label{sec:NNBO4RMPC}
Robust formulations of MPC aim to enable the use of this control scheme in applications where uncertainties are too significant to be ignored and control actions need to be picked while accounting for a range of future possible evolutions of the system instead of a single prediction. Incorporating such complexity into the underlying optimisation problem of MPC results in much higher dimensionality and computational cost compared to non-robust schemes, thereby making the use of RMPC in real time even more difficult than LMPC and NMPC. For these same reasons, the incentives to approximate RMPC with NNBO are even higher. In this section we review existing methods to generate such approximations without guarantees, with probabilistic guarantees, and strict guarantees. Table 3 presents a short description of all reviewed methods along with performance metrics on case studies.

\begin{table}[H]
\caption{Summary of methods and frameworks for generating approximate RMPC controllers that use NNBO.}
\label{tab:NNBO4RMPCSummary}
\resizebox{\linewidth}{!}{%
\begin{tabular}{@{}ccp{6cm}p{4cm}cccccccc@{}}
\toprule

Ref & L & \multicolumn{1}{c}{Core Concept} & \multicolumn{1}{c}{Guarantees} & $n_x$ & $n_y$ & $n_u$ & $N_p$ & $N_c$ & $N_r$ & \multicolumn{1}{c}{Speed up} & Validation \\ 

\midrule
\cite{lucia2018deep} & 1a & Train a DNN via simple imitation learning with MSE loss to learn scenario RMPC & None & 7 & 1 & 3 & 15 & 15 & 1 & 2,800x & Sim \\
\cite{karg2019learning} & 1e & Train DNN to imitate msNMPC, use constraint back-off parameters during training data generation to improve likelihood of constraint satisfaction. & None & 3 & 3 & 1 & 80 & 80 & 1 & & Sim \\
\cite{bonzanini2021fast} & 1a & Use GPs to learn plant model time-varying mismatch and adapt scenario tree of scenario MPC online. Train DNN with MSE loss to imitate resulting controller. & None & 3 & 1 & 2 & 5 & 5 & 1 & 4.4x & Sim \\
\cite{karg2021reinforced} & 3c & Train a DNN with simple MSE loss to imitate real MPC controller, use the network to initialise an RL agent, improve RL agent via further training. & None & 7 & 1 & 3 & 20 & 20 &  1 &  & Sim \\
\cite{hertneck2018learning} & 2b & Apply HI to performance of DNN approximation over trajectories to ensure stability with high probability. & Probabilistic stability and constraint satisfaction based on HI & 2 & 2 & 1 & 180 & 180 &  & 200x & Sim \\
\cite{nubert2020safe} & 2b & Apply HI to performance of DNN approximation over trajectories to ensure stability with high probability. & Probabilistic stability and constraint satisfaction based on HI & 8 & 3 & 4 & & &  & 200x & Robot \\
\cite{karg2021probabilistic} & 2b & Finite families based method for obtaining probabilistic guarantees of controllers using non-binary performance indicators. & Probabilistic constraint satisfaction & 3 & 1 & 1 & 40 & 40 & 1 & & Sim \\
\cite{daosud2019efficient} & 3a & Train DNN to predict cost of msRMPC, use this prediction as terminal cost of reformulated controller with shorter horizon. & Stability and constraint satisfaction but not optimality & 7 & 4 & 2 & 15 & 15 & 1 & 2.3x & Sim \\
\cite{paulson2020approximate} & 3b & Use a projection operation to coerce the outputs of a DNN approximation into sets that provide theoretical guarantees. Formulate the projection operation as a pQP that can be solved offline. & Constraint satisfaction and under special conditions, input-to-state stability. & 2 & 1 & 1 & 10 & 10 & 2 & 400x & Sim \\

\bottomrule

\end{tabular}%
}
\end{table}

\subsection{Methods without guarantees}
\label{sec:NNBO4RMPC_none}

As for LMPC and NMPC, the simplest way to approximate RMPC with an NN, is to use imitation learning to learn the behaviour of the controller to a satisfactory level of performance. \cite{lucia2018deep} demonstrated that this approach is applicable to RMPC despite its higher complexity. Specifically, they used a DNN with $tanh$ activation functions to learn the behaviour of multi-stage RMPC (msRMPC). Through simulation examples they proved the success of the approach at approximating the equivalent robust nonlinear controller for an industrial polymerisation reactor. The approximate controller was 2,800 times faster than the exact controller and could successfully be executed at the required rate in real time, however, it should also be pointed out that the average constraint violation, despite remaining low, was one order of magnitude larger than that of the exact controller. Thus, the pitfall of the basic imitation learning setup remains the same as previously discussed: it is not suitable for use in safety-critical systems because it provides no guarantees that the resulting approximate controller will respect the constraints of the system. 

In follow up work, \cite{karg2019learning} presented a modified approach where instead of directly learning the behaviour of msRMPC, an NN was used to learn the behaviour of a slightly modified controller which contained a constraint back-off parameter that made the controller more conservative. Through simulation experiments on a small nonlinear plant with one input and three states, the authors showed that the modified imitation learning approach could be successfully used to create an approximate controller that respected the constraint to which the back-off was applied. In their example, the back-off parameter was applied to a scalar nonlinear constraint, no examples were provided for the case of multiple constraints on inputs and states. The revised method with back-off parameters is a reasonable compromise between controller conservativeness and safety. This is important because, even in systems that are not safety-critical, avoiding constraint violations may be desirable to prevent unexpected stops or mechanical failure. Nonetheless, it should be noted that the back-off strategy does not provide any formal guarantees of constraint satisfaction. By using it, constraint violation is less likely to occur and the authors outlined a method via which the probability of violation can be computed after the approximation is generated. Yet, it is important to note that the method itself does not inherently provide probabilistic or strict guarantees.

\cite{bonzanini2021fast} further embraced the idea that the lack of strict safety guarantees can be overlooked in favour of other valuable compromises such as enabling the use of real-time msRMPC controllers that use augmented models equipped with Gaussian Processes (GP) to learn and mitigate the effects of plant-model mismatch. The incentive for creating and using these augmented models is that the standard formulation of msRMPC assumes that the structure and distribution of uncertainties is bounded and known in advance. However, as the authors pointed out, uncertainty can be time-varying and some systems are just too hard to model using conventional approaches. In such cases GP regression can be used to learn the unmodeled dynamics or time-varying uncertainties of the system from data. The msRMPC formulation can then be modified to accommodate the GP-augmented models. Via simulation experiments on a system with significant plant-model mismatch, Bonzanini et al. showed that their learning-based reformulation of msRMPC outperformed the standard formulation but its computation cost remained too high for use in real time. To overcome this issue, they applied the basic imitation learning solution and trained a DNN to learn the behaviour of the reformulated controller. The resulting approximation enjoyed no safety guarantees, but its performance on closed-loop simulation examples matched that of the exact reformulated msRMPC controller at a quarter of the online computation cost. \cite{karg2021reinforced} proposed a different method that achieves a similar result as \cite{bonzanini2021fast}, instead of reformulating msRMPC to account for more complex models, they trained a DNN via imitation learning to approximate the standard form of msRMPC and then used the resulting network as an initialisation of RL. Further training of the initialised agent then took care of the problem of learning the unmodeled dynamics and time-varying uncertainties. Both of these studies \citep{bonzanini2021fast, karg2021reinforced}, are examples of accepting the lack of guarantees of the simplest NNBO approach to enable the use of complex controllers that build beyond classical MPC formulations.

\subsection{Methods with probabilistic guarantees}
\label{sec:NNBO4RMPC_probabilistic}

In this section, we present two frameworks that are designed to preserve certain theoretical properties of RMPC on a probabilistic basis. The first framework, proposed by \cite{hertneck2018learning}, was an extension of the pure regression approach which includes a validation step that once passed, equips the basic approximation with statistical guarantees on constraint satisfaction and stability. In short, the validation step tests a candidate DNN approximation of the original controller on several closed-loop trajectories which start and end in feasible regions of the system's state space, and which the original RMPC controller can solve without violating any constraints. For each trajectory in the test set, an indicator function is applied which outputs 1 if the closed-loop solution of the DNN is within a specified tolerance of the solution of the real controller, or 0 if it is not. The mean of the indicator outputs is equivalent to $\frac{1}{n} \sum_{i=1}^{n}\xi_i$ in HI \eqref{eq:hoeffdings}. It follows that, if the value of $E(\xi)$ and the RHS of \eqref{eq:hoeffdings} are predefined, then one can calculate $\epsilon$ and verify if the difference between $\frac{1}{n} \sum_{i=1}^{n}\xi_i$ and $\epsilon$ is greater or equal than the prescribed $E(\xi)$ for the current number of test trajectories. If the verification is not successful, then the number of test trajectories can be increased or the quality of the DNN approximation can be improved by modifying the DNN and re-training until the validation step succeeds. If a candidate solution passes the validation step, it carries the guarantee that with some predefined confidence, its solutions are stable and constraint compliant (as defined by the indicator function) with a probability of at least the selected $E(\xi)$. 

\cite{hertneck2018learning} applied this methodology to an RMPC formulation designed for robustness against bounded input disturbances. To preserve said robustness on the approximate controller, their choice of indicator function outputs 1 if the infinity norm of the input error between the real and approximate controllers over a closed loop trajectory is less or equal to the input disturbance bound of the original controller. Otherwise the indicator outputs 0. The effectiveness of the approach was demonstrated on a simulation study entailing the control of a continuously stirred tank reactor, in said study the performance of the approximate controller was found to be indistinguishable from that of the original policy while being over 200 times faster to evaluate. 

Later on, \cite{nubert2020safe} applied the same approach to a more complex version of RMPC designed for setpoint tracking of nonlinear systems with robustness against plant-model mismatch. In this case the choice of indicator function was adjusted to output 1 if the solutions of the approximate controller complied with the maximum admissible model mismatch for the approximation, and 0 otherwise. The effectiveness of the reformulated approach was demonstrated on setpoint tracking experiments with a real robot arm that was represented by a simplified model with 8 states, 4 inputs and 3 outputs for control purposes. \cite{nubert2020safe} reported difficulties in satisfying their HI validation criterion for high probability levels but the resulting approximate controller worked well in practice and could be executed 200 times faster than the original controller thereby allowing the use of this complex form of RMPC in real time. It should be noted that this is the only framework that has been used across multiple studies and proved to be scalable and adaptable to more complex controllers while retaining the same speed up factor. It is also one of the few frameworks that have been tested on experiments with real hardware instead of simulation.

\cite{karg2021probabilistic} proposed a different framework for obtaining statistical guarantees on approximations of msRMPC which is suitable for the case where the performance of a controller with respect to a scalar indicator depends on a set of hyper-parameters. In this framework a family of controller approximations is considered, where each member of the family corresponds to a controller with a particular combination of hyper parameters. Using concepts from order statistics and probabilistic robustness analysis, the authors came up with expressions from which the size of a test set can be computed as a function of the size of the family. For said test set the authors also derived expressions to calculate the number of tests that can be failed, the desired probability of indicator violation, and the desired confidence level of the probability of violation. Each controller in the finite family can be evaluated on this test set and if the number of failed tests is lower than the allowable number of failures, then the family member is guaranteed to carry the probabilistic properties that were prescribed. \cite{karg2021probabilistic} then showed how this framework could be applied to a family of DNN approximations where each family member was trained to approximate the behaviour of an msRMPC controller with a particular constraint tightening parameter. In this example, the scalar indicator function was set to the maximum constraint violation over a 60 $s$ simulation of the closed loop system and the target performance of the probabilistic validation was to allow no violation at all with a prescribed probability. By applying their framework the size of the test set was computed and the authors showed how the different members of the family met or failed the maximum number of failures requirement. A perk of this method is that it can be applied to any arbitrary controller, not just DNN approximations of msRMPC. However, for generating neural approximations of MPC, which is the focal point of this paper, the method requires creating a data set per family member being considered. That means evaluating expensive MPC policies on large data sets multiple times instead of once like in other approximation frameworks. Therefore, there seems to be no compelling reason to choose this approximation framework over the previously discussed methods \citep{hertneck2018learning, nubert2020safe} given that both can yield the same probabilistic guarantees.

\subsection{Methods with strict guarantees}
\label{sec:NNBO4RMPC_strict}

In this section we present methods for approximating RMPC with DNN in manners that preserve some of the theoretical properties of MPC such as the guarantees of constraint satisfaction and stability. \cite{daosud2019efficient} proposed an intuitive approach for preserving the stability and constraint satisfaction properties of msRMPC. Their method consists of training a DNN via imitation learning to predict the optimal cost of msRMPC for a given starting condition. They used this trained network as the terminal cost of a re-formulated msRMPC controller with a much shorter prediction horizon than the original. Shortening the prediction horizon greatly reduces the computational cost of optimisation but enables the re-formulated controller to retain the stability and constraint satisfaction properties of MPC because rigorous optimisation is still being conducted at each control interval. The authors demonstrated the effectiveness of the approach on a simulated semi-batch reactor problem consisting of a MIMO plant with dimensions as shown in Table \ref{tab:NNBO4RMPCSummary}. First, the original controller was implemented and simulated with a robust horizon of 1 step and a prediction horizon of 15 steps. Then, a DNN with $tanh$ activation functions was used to learn to predict the optimal cost of the solutions of this controller. Subsequently, the trained network was used as the terminal cost of a reformulated controller which had the same robust horizon as the original but a prediction horizon of only 4 steps. Through simulation experiments the authors verified that the new controller with the DNN terminal cost was stable, constraint compliant and ran 2.3 times faster on average than the original controller. The main advantages of this approach are its intuitive nature and that the computation time reductions should only improve as the length of the horizon of the original controller increases because the cost of evaluating the DNN would remain constant. The only disadvantage is that the optimality guarantees of MPC are not maintained. In their study, \cite{daosud2019efficient} reported a 5\% performance loss when using the approximate controller compared to the original in terms of their control goal which was to maximise the concentration of a reactant in the semi-batch reactor. This performance loss was attributed to the prediction error of the DNN and it should be noted that the worst case or average performance loss were not reported, so in practice this number could be much higher, particularly if the quality of the DNN predictions worsens for harder problems and longer horizons. Nevertheless, while some performance is bound to be lost and the computation time reduction is not immense, the intuitiveness and guarantees of this method make it worthy of more investigation into its scalability and applicability to other types of MPC such as NMPC.

The only other alternative with strict guarantees was proposed by \cite{paulson2020approximate} and relies on projection operations to preserve the constraint and stability guarantees of msRMPC. Their method consists of training a DNN to learn the mapping between states and control actions for standard msRMPC. Once the network is trained its outputs are projected via optimisation onto carefully selected sets that preserve certain properties of MPC. For preserving constraint satisfaction properties in the presence of uncertainties, the authors proposed projecting the DNN outputs onto the maximal RCI set. To preserve stability properties, the authors projected the results onto the maximal $\lambda$-contractive set, which is a set in state space where all states can be driven into a tighter region of the set by applying one control input in the absence of disturbances. With this approach one can have an approximate controller that is constraint compliant in the presence of disturbances and asymptotically stable in the absence of disturbances. Nonetheless, assuming an absence of disturbances defeats the purpose of using RMPC in the first place and while the use of the $\lambda$-contractive set can yield Input-to-State Stability (ISS) guarantees for disturbed cases, there are additional conditions that need to be satisfied in this case which are difficult and perhaps impossible to verify a priori. In fact, the requirements for the constraint satisfaction guarantees are also hard to meet because the RCI set cannot always be computed or approximated for complex system with large plants. The authors demonstrated their approach on a double integrator plant problem and reported a computation speed up of 400 times compared to the original controller when the optimisation based projection operation was solved offline as an mpQP. No speed up factors or results were reported for cases where the projection operation cannot be solved offline. The applicability of this method to more complex problems has not been studied but seems unlikely. 

As for the NMPC case, the options for generating approximations of RMPC with strict guarantees are very limited and rely on further optimisation or difficult to meet conditions. There are also few publications on the topic and only msRMPC has been considered so far. Other robust formulations like min-max RMPC and Tube MPC have not been considered. 

\section{Discussion}

We hereby make an effort to synthesise our findings and observations regarding the state of the research field on NNs for fast optimisation in MPC. We start with a discussion on the types of problems used in case studies. In this discussion we highlight the lack of rigorous benchmarks to validate frameworks and how this prevents meaningful and informative comparisons. We then summarise areas where more work is needed and reflect on promising avenues for future work. Lastly, we provide guidelines for choosing a framework for readers seeking to apply NNBO to their own application. 

\subsection{Problem types and the lack of benchmarks}

\begin{figure}[!ht]
    \centering
    \includegraphics[width=\linewidth]{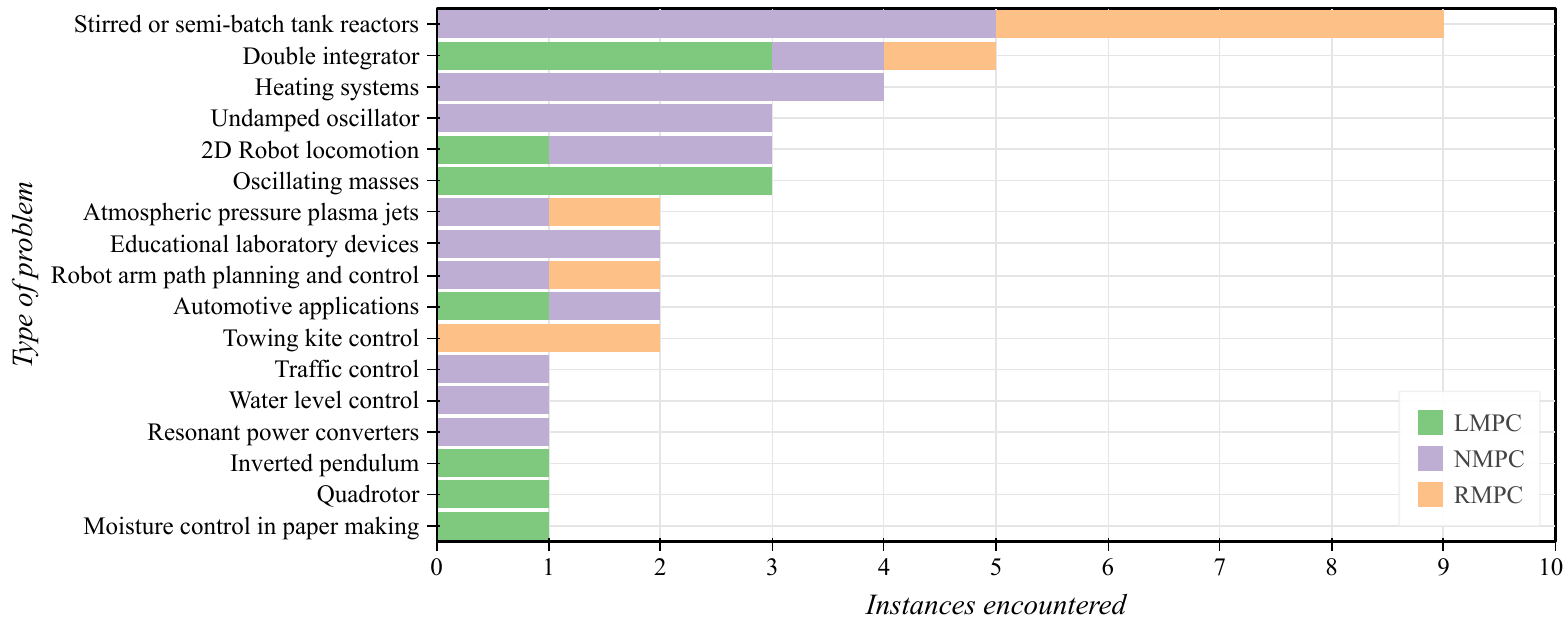}
    \caption{Number of instances of different types of problems considered in literature to demonstrate the effectiveness of NNBO approximation methods for LMPC, NMPC, and RMPC. }
    \label{fig:problemsBarChart}
\end{figure}

Figure \ref{fig:problemsBarChart} shows a bar chart with the types of problems considered in case studies to demonstrate NNBO frameworks. For each problem type, the chart indicates the number of instances encountered in the works reviewed. Some publications considered more than one type of problem. In such cases, we counted all types of problems used. On the contrary, a few studies used synthetic problems and did not state the physical meaning of the plant. For those cases we did not include the data in the figure.

From the graph it is clear that there are a few problem types that are significantly more popular, while other types have only been encountered once or twice. The trend we aim to highlight is the absence of well-established benchmarks for testing approximation frameworks. The stirred or semi-batch tank reactor problems are by far the most commonly encountered problem type in case studies. \cite{ren2022tutorial} made the same observation on their survey on NN modelling approaches for MPC. Specifically, they noted that chemical reaction problems are commonly encountered in case studies of machine learning-assisted optimal control because it is difficult to model such processes from first principles. In the works covered in this review, versions of the stirred reactor problem type are usually polymerisation reactions or pH control problems. Although the plants involved vary in complexity, most are nonlinear MIMO plants of medium size with 4 to 8 states and 2 to 3 inputs. The second most popular problem type is the double integrator, which is a SISO plant with 2 states. This problem type is not dissimilar from undamped oscillator problems. We encountered some publications \citep{paulson2020approximate, karg2020stability, pin2010approximate, pin2013approximate} that demonstrated their approach only on these problem types, which raises concerns about their scalability. Heating system problems are the third most popular type. The plants involved in these problems vary in size, ranging from medium to very large with hundreds of states. It is evident then that there is no standardised benchmark for showcasing the effectiveness of a given framework. 

Some research studies addressed the lack of benchmarks challenge by using multiple examples and more than one problem type to demonstrate an approach. For instance, \cite{fabiani2022reliably} used six versions of the oscillating masses problem with increasing dimensions and a fixed horizon length to demonstrate the scalability of their method. Similarly, \cite{chen2022large} considered three problem types and four case studies including the double integrator, quadrotor, and two instances of the oscillating masses problem. By demonstrating their approach on plants of different sizes and different nature, they not only proved the scalability of their method, but also its applicability beyond a specific problem type. Alternatively, \cite{li2022using} focused on a single plant but made efforts to demonstrate the scalability of their method in different ways. Most notably, they showed how the training time and performance of their approach varied as more training data was considered. Studies such as these allow to make important comparisons between approximation methods. In turn, these comparisons shed light on the information that a comprehensive benchmark should aim to measure.  

\begin{figure}[!ht]
    \centering
    \includegraphics[width=\linewidth]{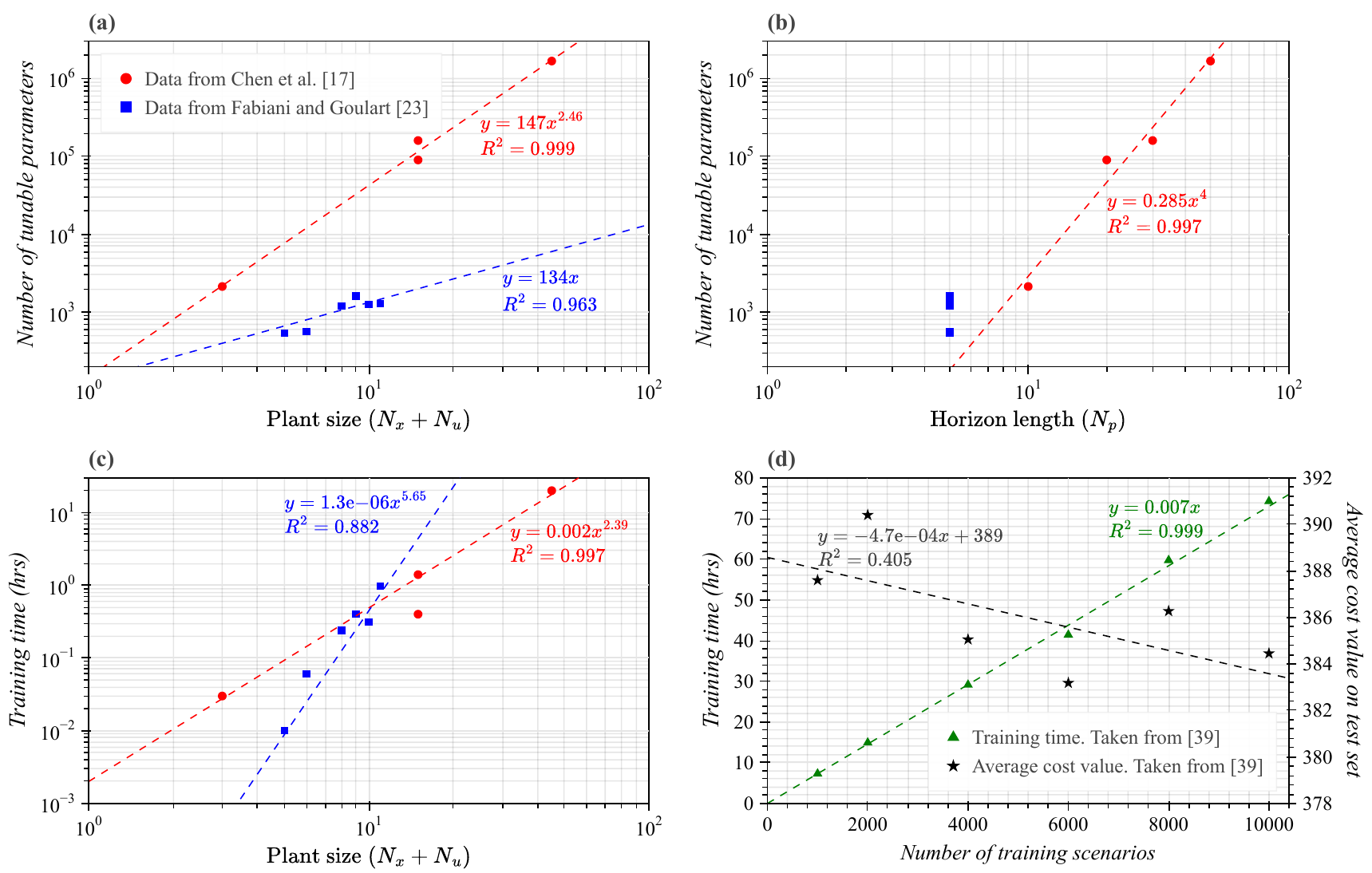}
    \caption{Examples of scalability estimates for frameworks by \cite{chen2022large}, \cite{fabiani2022reliably}, and \cite{li2022using}. (a) relationship between NN size and plant size. (b) relationship between NN size and prediction horizon length. (c) relationship between approximation training time and plant size. (d) relationship between training time and amount of training data, and approximation performance and amount of training data. }
    \label{fig:scalability}
\end{figure}

Figure \ref{fig:scalability} shows several relationships estimated using the data extracted from the case studies by \cite{fabiani2022reliably, chen2022large} and \cite{li2022using}. Figures \ref{fig:scalability}a -- c compare the frameworks by \cite{chen2022large} and \cite{fabiani2022reliably} which are both of the strict guarantees type for LMPC, but are otherwise very different in nature. The former consists of training a DNN that predicts all primal variables to hot-start an optimisation solver. The latter proposes MILPs to compute the values of properties that can be used to verify the stability and constraint compliance of controllers approximated via DNN ReLU regression. As shown in the plots, these differences in nature have profound impacts on the way that both methods scale for problems of different size. For example, the equations of the fitted functions in Figure \ref{fig:scalability}a show that the required network size in the method by \cite{chen2022large} scales exponentially with plant size. However, for the method by \cite{fabiani2022reliably} it scales linearly. This information is valuable because training large networks is likely to require more time and computational resources than for smaller networks. Note however, that because \cite{fabiani2022reliably} only considered one problem type of various sizes, it is not possible to say if this linear scaling holds in general. Figure \ref{fig:scalability}b shows how the required NN size varies with prediction horizon length. Since \cite{fabiani2022reliably} considered a fixed horizon length for all case studies, no comparisons can be made between the frameworks. For the \cite{chen2022large} framework, the plot clearly shows that the required network size grows even quicker with horizon length than it does with plant size. Knowing this would be particularly useful when seeking to apply the method to an existing controller. If the required horizon length is too long, the required network size could deem an approach inapplicable to the problem at hand. Lastly, Figure \ref{fig:scalability}c shows how training time varies with plant size for both approximation methods. It should be noted that \cite{fabiani2022reliably} reported the solution times required for the MILPs involved in their approach but did not make separate reports on the training time for the DNNs. Despite this, the available data indicates that training time for their method increases faster with plant size than it does for the method by \cite{chen2022large}. This information would be useful to know at the time of choosing an approximation framework for a new application. However, in this case the observed trends could be an artifact of the computational resources used in the respective studies. A fair comparison in training time would require performing both studies on the same hardware, and even then, differences in software implementations could skew the results.    

Perhaps the more meaningful information pertaining to training time, is how it affects the performance of the resulting approximation. Figure \ref{fig:scalability}d shows this relationship as reported by \cite{li2022using} in their case studies. Data in the plot was generated for a neural approximation and control problem of fixed size, meaning that the plant and prediction horizon of the original controller remained constant for all experiments. Only the amount of data used to train the network was varied. As expected, increasing the amount of training data increases the required training time, but as \cite{li2022using} noted, it also tends to increase performance. In the plot performance is expressed as ``average cost value". This metric is the mean of the objective function in the original MPC optimisation problem evaluated for every test point with the values produced by the NN approximation. Thus, better performance corresponds to a lower average cost. The linear regression of the average cost data in the plot confirms that performance does tend to increase with increasing amounts of training data. The gradient of the fit is small and its quality is low $(R^2=0.405)$ indicating that this effect is weak. Nevertheless, it is possible that not all approximation frameworks exhibit this relationship, or that they present different relationships with possibly stronger correlations. For practical applications knowing this information before applying a framework would be very useful. Another important point is that the framework by \cite{li2022using} does not involve generating training data by repeatedly solving MPC problems. Instead, the method is an optimise and train approach where the MPC problems are solved at the same time as the NN approximation is trained, all as part of a unified problem. The time involved in data generation for other approaches is non-negligible. For instance, \cite{chen2022large} reported up to $5.2\;hr$ for their largest case study despite using an efficient geometric walk approach. Therefore, in addition to measuring all aforementioned scaling properties, an adequate benchmarking tool should also aim to measure total approximation time, including both data generation and training time. 

Having discussed some of the comparisons that can be made when more rigorous testing is involved, we now make an effort to define what a truly informative benchmark should look like. First, the proposed benchmark should consider a wide range of problem types and dimensions for all versions of MPC, chosen with the intent of assessing scalability and applicability beyond specific cases. With such a suite of problems, the scaling relationships in Figures \ref{fig:scalability}a and \ref{fig:scalability}b could be defined for all frameworks. However, unlike some of the data presented in the figures, experiments should be conducted by varying only one variable at a time and conducting multiple experiments. For instance, the benchmark could measure how the required network size scales with plant size for horizons of fixed short, medium, and long lengths. Similarly, the scaling with horizon length could be measured for small, medium, and large plants. Such experiments should also involve a diverse combination of problem types and SISO and MIMO plants, so that the identified trends can be assumed to hold in general. In terms of computation time, each experiment should measure the time required to generate training data, training time, and the factor by which the resulting approximation speeds up the original controller. To facilitate fair comparisons between frameworks, these metrics should be recorded alongside detailed descriptions of the hardware and software used for the tests, including details of the implementation of the original MPC controller. Regarding performance, all tests should measure metrics that quantify the compromises involved by using the resulting approximation instead of the original MPC controller. \cite{li2022using} provided good examples of such metrics including the ``average cost value" reported in Figure \ref{fig:scalability}d, as well as, the maximum and average constraint violation encountered during tests. However, since average cost value is problem specific, it would make more sense to normalise this value by dividing the average cost by the average cost of the original MPC controller. For a perfect approximation of the controller, this performance index would be 1. For imperfect approximations, it would be greater than 1 and become larger as the quality of the approximation decreases. 

Overall, rigorous testing could contribute to a deeper understanding of the limitations of existing frameworks and to identifying areas where more research is needed. The development and maintenance of the benchmarking tool described here should be a joint, open-source, research effort. A collaborative approach would increase transparency and allow for ongoing adaptation of the problem selections to ensure their relevance over time. We also expect that the development of such a tool would bring new interest to the field and help develop more flexible and robust NNBO frameworks for MPC.

\subsection{Areas for future research}

We start our discussion of the future of the field by summarising the gaps that have become immediately apparent in reviewing all the works we have covered. This summary is a list of opportunities to make contributions in short to medium time frames. Regarding the development of new NNBO frameworks, there are no existing methods that address the general case of integer control variables. Existing studies have only treated the case of binary variables \citep{lohr2019mimicking, karg2018deep, lohr2020machine} and in all cases only a single integer variable was considered. For NMPC only a few frameworks in \citep{parisini1998nonlinear, pin2010approximate, pin2013approximate,ahmed1998neural} offer strict guarantees without relying on further optimisation or projection operations. However, as summarised in Table \ref{tab:NNBO4NMPCSummary} these frameworks can only preserve select guarantees, they have only been demonstrated on small problems, and their achieved speedup factors has not been reported. For RMPC all frameworks with strict guarantees also rely on further optimisation or projection \citep{daosud2019efficient, paulson2020approximate}. Moreover, all research concerning strict guarantees and no guarantees has focused on msRMPC, while min-max RMPC and Tube RMPC remain unaddressed. Another topic that remains unexplored is the use of more advanced neural architectures to generate approximations. The studies we covered mostly use shallow networks or multi-layer perceptrons with standard neurons and conventional activation functions. Only a few studies explored the use of more involved ML tools, such as LSTM units \citep{kumar2018deep} or autoencoders \citep{nurbayeva2022deep}. Many promising ML tools remain unused in the field, particularly those that can exploit temporal or sequential data and prior knowledge of the system. For instance, the use of transformers, which are best known for their use in natural language processing, has already been demonstrated in MPC for the purpose of modelling and their use in replacing the optimiser has also been suggested \citep{park2023simultaneous}. Another promising option is the use of Physics Informed NNs (PINNs), which are designed to respect the underlying physics of the process they model. This type of network has been recently adapted for use in MPC as a plant replacement \citep{antonelo2021physics}. However, the use of PINNs as part of an NNBO framework has not been explored. Lastly, other application oriented contributions could also be made along the lines of developing open source libraries that implement approximation frameworks for general purpose use or other NNBO tools. As already discussed, the field could greatly benefit from the development of good benchmarking tools.  

Beyond the evident gaps, we believe that future breakthroughs in the field lie in the development of ties between control and learning theory. \cite{karg2020stability} and \cite{fabiani2022reliably} pioneered this approach for LMPC. Their combined work resulted in a framework with strict optimality, stability, and recursive feasibility guarantees, which does not rely on post processing operations like projection or further optimisation. Instead, the result is a NN that can be used as a drop-in replacement for the original controller, just like for most pure regression approaches. Extending these results to NMPC and RMPC will require more theoretical work to create stronger links between control and learning theory. The lack of such ties has been noted more often in learning-based control research where it is especially difficult to guarantee the safety and robustness of control algorithms. \cite{shi2023reliable} treated this problem extensively noting that both the learning and control fields can greatly benefit from the development of unified frameworks that ease translation and encourage new discoveries. It is possible then that by studying efforts in translating control principles to learning principles, one might find ways to formulate new MPC approximation methods with strict guarantees. For instance, \cite{lin2021perturbation} worked on proving the dynamic regret properties of LVMPC, thereby creating new translation tools between learning and control. None of the works reviewed here use or exploit the concept of regret. Alternatively, \cite{shi2019neural} showed how spectral normalisation can be used to constrain the Lipschitz constant of a DNN and train nonlinear DNN-based controllers with strict stability guarantees. The resulting controller was purely learning-based, but it might be possible to adapt this novel training method for the purpose of approximating MPC with strict guarantees. 

\subsection{Selection guide for new applications}

\begin{figure}[!ht]
    \centering
    \includegraphics[width=10cm]{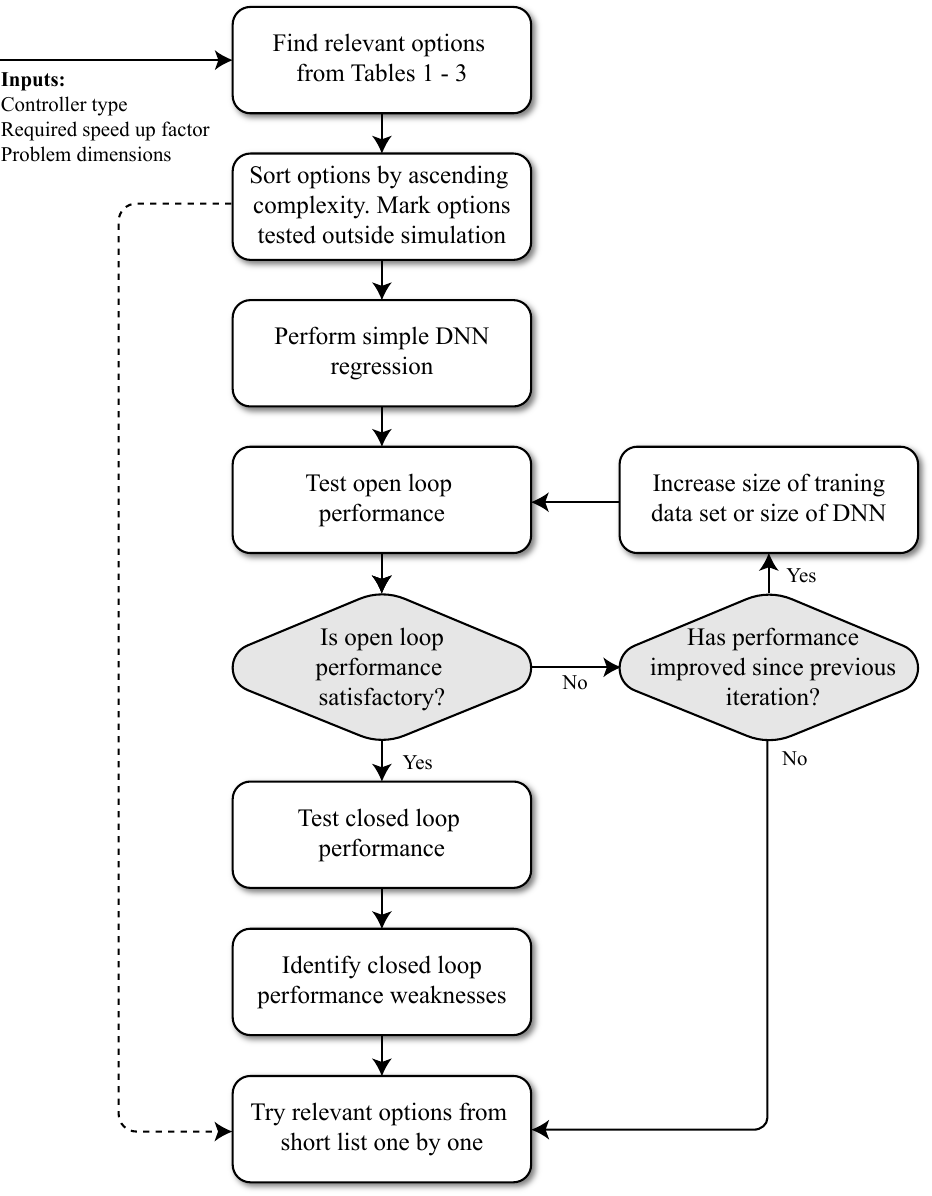}
    \caption{Summary of heuristic for selecting an NNBO framework to accelerate an existing MPC controller.}
    \label{fig:SelectionProcedureDiagram}
\end{figure}

The motivation for conducting this review originated from our personal experience in seeking an NNBO framework for a specific application. Drawing from this experience, we put forth some straightforward guidelines, summarised in Figure \ref{fig:SelectionProcedureDiagram}, for readers attempting to do the same.
The readers will likely undertake this task with the intention of enhancing the performance of an existing controller in terms of computational load. Where that is not the case, the first step is to define the type of controller that needs to be approximated e.g. LMPC, NMPC, or RMPC. Next, one must determine the factor by which the controller needs to be accelerated and take notes of the dimensions of the plant and the required horizon lengths. With this information, one can use Tables \ref{tab:NNBO4LMPCSummary} to \ref{tab:NNBO4RMPCSummary} to try to find a study that considers a plant of similar or larger dimensions and achieves a speed up factor close to what is required. When multiple options are available, it is important to consider all the available choices. The shortlist should be sorted by order of ascending complexity and options that have been shown to work outside of simulation should be prioritised. Next, the reader should consider the approximation methods involved in the frameworks and the types of guarantees that they yield. If the application in mind is safety critical, a framework with probabilistic or strict guarantees will likely be desirable, if any exist in the shortlisted options. The reader should then test the DNN imitation learning pure regression approach on their specific problem, which in all likelihood, will be one of the shortlisted options. This method will be the fastest approach for generating an approximation and it will provide a good estimate of the minimum DNN size required to imitate the controller. In addition, if the list of candidate frameworks is empty, because no matches were found for the problem size or required speed up factor, the pure regression method will provide a good idea of the feasibility of performing the approximation and achieving the desired speedup. 

After applying pure regression, one should test the quality of the resulting DNN approximation on open and closed loop control. Comparing the open loop performance of the approximation to that of the real controller will indicate if the approximation is a good fit or not. Closed loop control tests should include various tasks such as maintaining a steady state with and without disturbances, tracking step changes in reference, and using the controller as intended. These tests will reveal more insights such as the stability properties of the approximation, the likelihood of state and output constraint violations during operation, and the effect of cumulative error on the results achieved compared to the original controller. Once this information is collected, one can make a better informed decision on what options from the shortlist to try next. For instance, if the simple approximation is relatively stable but suffers from significant steady state errors, one might consider an option that addresses steady state error compensation. If closed loop performance is unstable, the search can be directed towards frameworks featuring MPC inspired loss function, data generation on closed loop trajectories, and frameworks designed to preserve guarantees. When choosing from these more advanced options the reader should consider if the requirements of the selected option can be fulfilled. Some options may require performing computations which may be known to be intractable for the problem at hand. For instance, computing the RCI or RAI sets for large nonlinear plants, or perhaps solving large MILPs or mpQPs. In such cases, one may prefer to prioritise testing other options from the shortlist.

\section{Conclusion}

In this survey we have presented, explained and compared in detail methods and frameworks for generating approximations of MPC controllers with NNBO. From the works reviewed we have proposed a taxonomy to classify existing and future methods. This taxonomy includes labels that indicate the type of MPC a framework is applicable to, the type of guarantees it offers, and the manner in which the approximation is generated. By collecting data from case studies in reviewed publications, we have provided tabulated summaries of known achievable performance that can help the reader choose an approximation framework for their own application. Moreover, through the process of reviewing and organising existing knowledge on the subject, we have uncovered several areas of research that remain unaddressed, a noticeable absence of rigorous benchmarks, and the research directions in which future breakthroughs might lie. From an application oriented point of view, the development of well defined benchmarks is instrumental in helping the field advance. From a theory point of view, future breakthroughs lie in the development of ties between control and learning theory.  

\bibliographystyle{elsarticle-harv} 
\bibliography{references}

\begin{thebibliography}{77}
\expandafter\ifx\csname natexlab\endcsname\relax\def\natexlab#1{#1}\fi
\providecommand{\url}[1]{\texttt{#1}}
\providecommand{\href}[2]{#2}
\providecommand{\path}[1]{#1}
\providecommand{\DOIprefix}{doi:}
\providecommand{\ArXivprefix}{arXiv:}
\providecommand{\URLprefix}{URL: }
\providecommand{\Pubmedprefix}{pmid:}
\providecommand{\doi}[1]{\href{http://dx.doi.org/#1}{\path{#1}}}
\providecommand{\Pubmed}[1]{\href{pmid:#1}{\path{#1}}}
\providecommand{\bibinfo}[2]{#2}
\ifx\xfnm\relax \def\xfnm[#1]{\unskip,\space#1}\fi
\bibitem[{Ahmed and Al-Dajani(1998)}]{ahmed1998neural}
\bibinfo{author}{Ahmed, M.}, \bibinfo{author}{Al-Dajani, M.}, \bibinfo{year}{1998}.
\newblock \bibinfo{title}{Neural regulator design}.
\newblock \bibinfo{journal}{Neural Networks} \bibinfo{volume}{11}, \bibinfo{pages}{1695--1709}.
\bibitem[{{\AA}kesson and Toivonen(2006)}]{aakesson2006neural}
\bibinfo{author}{{\AA}kesson, B.M.}, \bibinfo{author}{Toivonen, H.T.}, \bibinfo{year}{2006}.
\newblock \bibinfo{title}{A neural network model predictive controller}.
\newblock \bibinfo{journal}{Journal of Process Control} \bibinfo{volume}{16}, \bibinfo{pages}{937--946}.
\bibitem[{{\AA}kesson et~al.(2005){\AA}kesson, Toivonen, Waller and Nystr{\"o}m}]{aakesson2005neural}
\bibinfo{author}{{\AA}kesson, B.M.}, \bibinfo{author}{Toivonen, H.T.}, \bibinfo{author}{Waller, J.B.}, \bibinfo{author}{Nystr{\"o}m, R.H.}, \bibinfo{year}{2005}.
\newblock \bibinfo{title}{Neural network approximation of a nonlinear model predictive controller applied to a ph neutralization process}.
\newblock \bibinfo{journal}{Computers \& chemical engineering} \bibinfo{volume}{29}, \bibinfo{pages}{323--335}.
\bibitem[{Alessio and Bemporad(2009)}]{alessio2009survey}
\bibinfo{author}{Alessio, A.}, \bibinfo{author}{Bemporad, A.}, \bibinfo{year}{2009}.
\newblock \bibinfo{title}{A survey on explicit model predictive control}.
\newblock \bibinfo{journal}{Nonlinear Model Predictive Control: Towards New Challenging Applications} , \bibinfo{pages}{345--369}.
\bibitem[{Antonelo et~al.(2021)Antonelo, Camponogara, Seman, de~Souza, Jordanou and Hubner}]{antonelo2021physics}
\bibinfo{author}{Antonelo, E.A.}, \bibinfo{author}{Camponogara, E.}, \bibinfo{author}{Seman, L.O.}, \bibinfo{author}{de~Souza, E.R.}, \bibinfo{author}{Jordanou, J.P.}, \bibinfo{author}{Hubner, J.F.}, \bibinfo{year}{2021}.
\newblock \bibinfo{title}{Physics-informed neural nets for control of dynamical systems}.
\newblock \bibinfo{journal}{arXiv preprint arXiv:2104.02556} .
\bibitem[{Barron(1993)}]{barron1993universal}
\bibinfo{author}{Barron, A.R.}, \bibinfo{year}{1993}.
\newblock \bibinfo{title}{Universal approximation bounds for superpositions of a sigmoidal function}.
\newblock \bibinfo{journal}{IEEE Trans. Inf. Theor.} \bibinfo{volume}{39}, \bibinfo{pages}{930–945}.
\newblock \URLprefix \url{https://doi.org/10.1109/18.256500}, \DOIprefix\doi{10.1109/18.256500}.
\bibitem[{Bemporad and Filippi(2006)}]{bemporad2006algorithm}
\bibinfo{author}{Bemporad, A.}, \bibinfo{author}{Filippi, C.}, \bibinfo{year}{2006}.
\newblock \bibinfo{title}{An algorithm for approximate multiparametric convex programming}.
\newblock \bibinfo{journal}{Computational optimization and applications} \bibinfo{volume}{35}, \bibinfo{pages}{87--108}.
\bibitem[{Bemporad et~al.(2002)Bemporad, Morari, Dua and Pistikopoulos}]{bemporad2002explicit}
\bibinfo{author}{Bemporad, A.}, \bibinfo{author}{Morari, M.}, \bibinfo{author}{Dua, V.}, \bibinfo{author}{Pistikopoulos, E.N.}, \bibinfo{year}{2002}.
\newblock \bibinfo{title}{The explicit linear quadratic regulator for constrained systems}.
\newblock \bibinfo{journal}{Automatica} \bibinfo{volume}{38}, \bibinfo{pages}{3--20}.
\bibitem[{Bemporad et~al.(2011)Bemporad, Oliveri, Poggi and Storace}]{bemporad2011ultra}
\bibinfo{author}{Bemporad, A.}, \bibinfo{author}{Oliveri, A.}, \bibinfo{author}{Poggi, T.}, \bibinfo{author}{Storace, M.}, \bibinfo{year}{2011}.
\newblock \bibinfo{title}{Ultra-fast stabilizing model predictive control via canonical piecewise affine approximations}.
\newblock \bibinfo{journal}{IEEE Transactions on Automatic Control} \bibinfo{volume}{56}, \bibinfo{pages}{2883--2897}.
\bibitem[{Bonzanini et~al.(2020)Bonzanini, Paulson, Graves and Mesbah}]{bonzanini2020toward}
\bibinfo{author}{Bonzanini, A.D.}, \bibinfo{author}{Paulson, J.A.}, \bibinfo{author}{Graves, D.B.}, \bibinfo{author}{Mesbah, A.}, \bibinfo{year}{2020}.
\newblock \bibinfo{title}{Toward safe dose delivery in plasma medicine using projected neural network-based fast approximate nmpc}.
\newblock \bibinfo{journal}{IFAC-PapersOnLine} \bibinfo{volume}{53}, \bibinfo{pages}{5279--5285}.
\bibitem[{Bonzanini et~al.(2021)Bonzanini, Paulson, Makrygiorgos and Mesbah}]{bonzanini2021fast}
\bibinfo{author}{Bonzanini, A.D.}, \bibinfo{author}{Paulson, J.A.}, \bibinfo{author}{Makrygiorgos, G.}, \bibinfo{author}{Mesbah, A.}, \bibinfo{year}{2021}.
\newblock \bibinfo{title}{Fast approximate learning-based multistage nonlinear model predictive control using gaussian processes and deep neural networks}.
\newblock \bibinfo{journal}{Computers \& Chemical Engineering} \bibinfo{volume}{145}, \bibinfo{pages}{107174}.
\bibitem[{Cao and Gopaluni(2020)}]{cao2020deep}
\bibinfo{author}{Cao, Y.}, \bibinfo{author}{Gopaluni, R.B.}, \bibinfo{year}{2020}.
\newblock \bibinfo{title}{Deep neural network approximation of nonlinear model predictive control}.
\newblock \bibinfo{journal}{IFAC-PapersOnLine} \bibinfo{volume}{53}, \bibinfo{pages}{11319--11324}.
\bibitem[{Cavagnari et~al.(1999)Cavagnari, Magni and Scattolini}]{cavagnari1999neural}
\bibinfo{author}{Cavagnari, L.}, \bibinfo{author}{Magni, L.}, \bibinfo{author}{Scattolini, R.}, \bibinfo{year}{1999}.
\newblock \bibinfo{title}{Neural network implementation of nonlinear receding-horizon control}.
\newblock \bibinfo{journal}{Neural computing \& applications} \bibinfo{volume}{8}, \bibinfo{pages}{86--92}.
\bibitem[{Chakrabarty et~al.(2016)Chakrabarty, Dinh, Corless, Rundell, {\.Z}ak and Buzzard}]{chakrabarty2016support}
\bibinfo{author}{Chakrabarty, A.}, \bibinfo{author}{Dinh, V.}, \bibinfo{author}{Corless, M.J.}, \bibinfo{author}{Rundell, A.E.}, \bibinfo{author}{{\.Z}ak, S.H.}, \bibinfo{author}{Buzzard, G.T.}, \bibinfo{year}{2016}.
\newblock \bibinfo{title}{Support vector machine informed explicit nonlinear model predictive control using low-discrepancy sequences}.
\newblock \bibinfo{journal}{IEEE Transactions on Automatic Control} \bibinfo{volume}{62}, \bibinfo{pages}{135--148}.
\bibitem[{Chan et~al.(2021)Chan, Paulson and Mesbah}]{chan2021deep}
\bibinfo{author}{Chan, K.J.}, \bibinfo{author}{Paulson, J.A.}, \bibinfo{author}{Mesbah, A.}, \bibinfo{year}{2021}.
\newblock \bibinfo{title}{Deep learning-based approximate nonlinear model predictive control with offset-free tracking for embedded applications}, in: \bibinfo{booktitle}{2021 American Control Conference (ACC)}, \bibinfo{organization}{IEEE}. pp. \bibinfo{pages}{3475--3481}.
\bibitem[{Chen et~al.(2018)Chen, Saulnier, Atanasov, Lee, Kumar, Pappas and Morari}]{chen2018approximating}
\bibinfo{author}{Chen, S.}, \bibinfo{author}{Saulnier, K.}, \bibinfo{author}{Atanasov, N.}, \bibinfo{author}{Lee, D.D.}, \bibinfo{author}{Kumar, V.}, \bibinfo{author}{Pappas, G.J.}, \bibinfo{author}{Morari, M.}, \bibinfo{year}{2018}.
\newblock \bibinfo{title}{Approximating explicit model predictive control using constrained neural networks}, in: \bibinfo{booktitle}{2018 Annual American control conference (ACC)}, \bibinfo{organization}{IEEE}. pp. \bibinfo{pages}{1520--1527}.
\bibitem[{Chen et~al.(2022)Chen, Wang, Atanasov, Kumar and Morari}]{chen2022large}
\bibinfo{author}{Chen, S.W.}, \bibinfo{author}{Wang, T.}, \bibinfo{author}{Atanasov, N.}, \bibinfo{author}{Kumar, V.}, \bibinfo{author}{Morari, M.}, \bibinfo{year}{2022}.
\newblock \bibinfo{title}{Large scale model predictive control with neural networks and primal active sets}.
\newblock \bibinfo{journal}{Automatica} \bibinfo{volume}{135}, \bibinfo{pages}{109947}.
\bibitem[{Csek{\H{o}} et~al.(2015)Csek{\H{o}}, Kvasnica and Lantos}]{csekHo2015explicit}
\bibinfo{author}{Csek{\H{o}}, L.H.}, \bibinfo{author}{Kvasnica, M.}, \bibinfo{author}{Lantos, B.}, \bibinfo{year}{2015}.
\newblock \bibinfo{title}{Explicit mpc-based rbf neural network controller design with discrete-time actual kalman filter for semiactive suspension}.
\newblock \bibinfo{journal}{IEEE Transactions on Control Systems Technology} \bibinfo{volume}{23}, \bibinfo{pages}{1736--1753}.
\bibitem[{Daosud et~al.(2019)Daosud, Kittisupakorn, Fikar, Lucia and Paulen}]{daosud2019efficient}
\bibinfo{author}{Daosud, W.}, \bibinfo{author}{Kittisupakorn, P.}, \bibinfo{author}{Fikar, M.}, \bibinfo{author}{Lucia, S.}, \bibinfo{author}{Paulen, R.}, \bibinfo{year}{2019}.
\newblock \bibinfo{title}{Efficient robust nonlinear model predictive control via approximate multi-stage programming: A neural networks based approach}, in: \bibinfo{booktitle}{Computer Aided Chemical Engineering}. \bibinfo{publisher}{Elsevier}. volume~\bibinfo{volume}{46}, pp. \bibinfo{pages}{1261--1266}.
\bibitem[{Diehl and Bjornberg(2004)}]{diehl2004robust}
\bibinfo{author}{Diehl, M.}, \bibinfo{author}{Bjornberg, J.}, \bibinfo{year}{2004}.
\newblock \bibinfo{title}{Robust dynamic programming for min-max model predictive control of constrained uncertain systems}.
\newblock \bibinfo{journal}{IEEE Transactions on Automatic Control} \bibinfo{volume}{49}, \bibinfo{pages}{2253--2257}.
\bibitem[{Drgo{\v{n}}a et~al.(2022)Drgo{\v{n}}a, Ki{\v{s}}, Tuor, Vrabie and Klau{\v{c}}o}]{drgovna2022differentiable}
\bibinfo{author}{Drgo{\v{n}}a, J.}, \bibinfo{author}{Ki{\v{s}}, K.}, \bibinfo{author}{Tuor, A.}, \bibinfo{author}{Vrabie, D.}, \bibinfo{author}{Klau{\v{c}}o, M.}, \bibinfo{year}{2022}.
\newblock \bibinfo{title}{Differentiable predictive control: Deep learning alternative to explicit model predictive control for unknown nonlinear systems}.
\newblock \bibinfo{journal}{Journal of Process Control} \bibinfo{volume}{116}, \bibinfo{pages}{80--92}.
\bibitem[{Drgo{\v{n}}a et~al.(2018)Drgo{\v{n}}a, Picard, Kvasnica and Helsen}]{drgovna2018approximate}
\bibinfo{author}{Drgo{\v{n}}a, J.}, \bibinfo{author}{Picard, D.}, \bibinfo{author}{Kvasnica, M.}, \bibinfo{author}{Helsen, L.}, \bibinfo{year}{2018}.
\newblock \bibinfo{title}{Approximate model predictive building control via machine learning}.
\newblock \bibinfo{journal}{Applied Energy} \bibinfo{volume}{218}, \bibinfo{pages}{199--216}.
\bibitem[{Fabiani and Goulart(2022)}]{fabiani2022reliably}
\bibinfo{author}{Fabiani, F.}, \bibinfo{author}{Goulart, P.J.}, \bibinfo{year}{2022}.
\newblock \bibinfo{title}{Reliably-stabilizing piecewise-affine neural network controllers}.
\newblock \bibinfo{journal}{IEEE Transactions on Automatic Control} , \bibinfo{pages}{1--15}\DOIprefix\doi{10.1109/TAC.2022.3216978}.
\bibitem[{Ferreau et~al.(2017)Ferreau, Alm{\'e}r, Verschueren, Diehl, Frick, Domahidi, Jerez, Stathopoulos and Jones}]{ferreau2017embedded}
\bibinfo{author}{Ferreau, H.J.}, \bibinfo{author}{Alm{\'e}r, S.}, \bibinfo{author}{Verschueren, R.}, \bibinfo{author}{Diehl, M.}, \bibinfo{author}{Frick, D.}, \bibinfo{author}{Domahidi, A.}, \bibinfo{author}{Jerez, J.}, \bibinfo{author}{Stathopoulos, G.}, \bibinfo{author}{Jones, C.}, \bibinfo{year}{2017}.
\newblock \bibinfo{title}{Embedded optimization methods for industrial automatic control}.
\newblock \bibinfo{journal}{IFAC-PapersOnLine} \bibinfo{volume}{50}, \bibinfo{pages}{13194--13209}.
\bibitem[{Ferreau et~al.(2014)Ferreau, Kirches, Potschka, Bock and Diehl}]{ferreau2014qpoases}
\bibinfo{author}{Ferreau, H.J.}, \bibinfo{author}{Kirches, C.}, \bibinfo{author}{Potschka, A.}, \bibinfo{author}{Bock, H.G.}, \bibinfo{author}{Diehl, M.}, \bibinfo{year}{2014}.
\newblock \bibinfo{title}{qpoases: A parametric active-set algorithm for quadratic programming}.
\newblock \bibinfo{journal}{Mathematical Programming Computation} \bibinfo{volume}{6}, \bibinfo{pages}{327--363}.
\bibitem[{Hertneck et~al.(2018)Hertneck, K{\"o}hler, Trimpe and Allg{\"o}wer}]{hertneck2018learning}
\bibinfo{author}{Hertneck, M.}, \bibinfo{author}{K{\"o}hler, J.}, \bibinfo{author}{Trimpe, S.}, \bibinfo{author}{Allg{\"o}wer, F.}, \bibinfo{year}{2018}.
\newblock \bibinfo{title}{Learning an approximate model predictive controller with guarantees}.
\newblock \bibinfo{journal}{IEEE Control Systems Letters} \bibinfo{volume}{2}, \bibinfo{pages}{543--548}.
\bibitem[{Hewing et~al.(2020)Hewing, Wabersich, Menner and Zeilinger}]{hewing2020learning}
\bibinfo{author}{Hewing, L.}, \bibinfo{author}{Wabersich, K.P.}, \bibinfo{author}{Menner, M.}, \bibinfo{author}{Zeilinger, M.N.}, \bibinfo{year}{2020}.
\newblock \bibinfo{title}{Learning-based model predictive control: Toward safe learning in control}.
\newblock \bibinfo{journal}{Annual Review of Control, Robotics, and Autonomous Systems} \bibinfo{volume}{3}, \bibinfo{pages}{269--296}.
\bibitem[{Jin et~al.(2019)Jin, Li, Hu and Liu}]{jin2019survey}
\bibinfo{author}{Jin, L.}, \bibinfo{author}{Li, S.}, \bibinfo{author}{Hu, B.}, \bibinfo{author}{Liu, M.}, \bibinfo{year}{2019}.
\newblock \bibinfo{title}{A survey on projection neural networks and their applications}.
\newblock \bibinfo{journal}{Applied Soft Computing} \bibinfo{volume}{76}, \bibinfo{pages}{533--544}.
\bibitem[{Johansen(2004)}]{johansen2004approximate}
\bibinfo{author}{Johansen, T.A.}, \bibinfo{year}{2004}.
\newblock \bibinfo{title}{Approximate explicit receding horizon control of constrained nonlinear systems}.
\newblock \bibinfo{journal}{Automatica} \bibinfo{volume}{40}, \bibinfo{pages}{293--300}.
\bibitem[{Karg et~al.(2021)Karg, Alamo and Lucia}]{karg2021probabilistic}
\bibinfo{author}{Karg, B.}, \bibinfo{author}{Alamo, T.}, \bibinfo{author}{Lucia, S.}, \bibinfo{year}{2021}.
\newblock \bibinfo{title}{Probabilistic performance validation of deep learning-based robust nmpc controllers}.
\newblock \bibinfo{journal}{International Journal of Robust and Nonlinear Control} \bibinfo{volume}{31}, \bibinfo{pages}{8855--8876}.
\bibitem[{Karg and Lucia(2018)}]{karg2018deep}
\bibinfo{author}{Karg, B.}, \bibinfo{author}{Lucia, S.}, \bibinfo{year}{2018}.
\newblock \bibinfo{title}{Deep learning-based embedded mixed-integer model predictive control}, in: \bibinfo{booktitle}{2018 european control conference (ecc)}, \bibinfo{organization}{IEEE}. pp. \bibinfo{pages}{2075--2080}.
\bibitem[{Karg and Lucia(2019)}]{karg2019learning}
\bibinfo{author}{Karg, B.}, \bibinfo{author}{Lucia, S.}, \bibinfo{year}{2019}.
\newblock \bibinfo{title}{Learning-based approximation of robust nonlinear predictive control with state estimation applied to a towing kite}, in: \bibinfo{booktitle}{2019 18th European Control Conference (ECC)}, \bibinfo{organization}{IEEE}. pp. \bibinfo{pages}{16--22}.
\bibitem[{Karg and Lucia(2020a)}]{karg2020efficient}
\bibinfo{author}{Karg, B.}, \bibinfo{author}{Lucia, S.}, \bibinfo{year}{2020}a.
\newblock \bibinfo{title}{Efficient representation and approximation of model predictive control laws via deep learning}.
\newblock \bibinfo{journal}{IEEE Transactions on Cybernetics} \bibinfo{volume}{50}, \bibinfo{pages}{3866--3878}.
\bibitem[{Karg and Lucia(2020b)}]{karg2020stability}
\bibinfo{author}{Karg, B.}, \bibinfo{author}{Lucia, S.}, \bibinfo{year}{2020}b.
\newblock \bibinfo{title}{Stability and feasibility of neural network-based controllers via output range analysis}, in: \bibinfo{booktitle}{2020 59th IEEE Conference on Decision and Control (CDC)}, \bibinfo{organization}{IEEE}. pp. \bibinfo{pages}{4947--4954}.
\bibitem[{Karg and Lucia(2021)}]{karg2021reinforced}
\bibinfo{author}{Karg, B.}, \bibinfo{author}{Lucia, S.}, \bibinfo{year}{2021}.
\newblock \bibinfo{title}{Reinforced approximate robust nonlinear model predictive control}, in: \bibinfo{booktitle}{2021 23rd International Conference on Process Control (PC)}, \bibinfo{organization}{IEEE}. pp. \bibinfo{pages}{149--156}.
\bibitem[{Kingma and Ba(2014)}]{kingma2014adam}
\bibinfo{author}{Kingma, D.P.}, \bibinfo{author}{Ba, J.}, \bibinfo{year}{2014}.
\newblock \bibinfo{title}{Adam: A method for stochastic optimization}.
\newblock \bibinfo{journal}{arXiv preprint arXiv:1412.6980} .
\bibitem[{Kouvaritakis and Cannon(2015)}]{kouvaritakis2015model}
\bibinfo{author}{Kouvaritakis, B.}, \bibinfo{author}{Cannon, M.}, \bibinfo{year}{2015}.
\newblock \bibinfo{title}{Model Predictive Control: Classical, Robust and Stochastic}.
\newblock \bibinfo{publisher}{Springer}.
\bibitem[{Kumar et~al.(2018)Kumar, Tulsyan, Gopaluni and Loewen}]{kumar2018deep}
\bibinfo{author}{Kumar, S.S.P.}, \bibinfo{author}{Tulsyan, A.}, \bibinfo{author}{Gopaluni, B.}, \bibinfo{author}{Loewen, P.}, \bibinfo{year}{2018}.
\newblock \bibinfo{title}{A deep learning architecture for predictive control}.
\newblock \bibinfo{journal}{IFAC-PapersOnLine} \bibinfo{volume}{51}, \bibinfo{pages}{512--517}.
\bibitem[{Li et~al.(2022)Li, Hua and Cao}]{li2022using}
\bibinfo{author}{Li, Y.}, \bibinfo{author}{Hua, K.}, \bibinfo{author}{Cao, Y.}, \bibinfo{year}{2022}.
\newblock \bibinfo{title}{Using stochastic programming to train neural network approximation of nonlinear mpc laws}.
\newblock \bibinfo{journal}{Automatica} \bibinfo{volume}{146}, \bibinfo{pages}{110665}.
\bibitem[{Lin et~al.(2021)Lin, Hu, Shi, Sun, Qu and Wierman}]{lin2021perturbation}
\bibinfo{author}{Lin, Y.}, \bibinfo{author}{Hu, Y.}, \bibinfo{author}{Shi, G.}, \bibinfo{author}{Sun, H.}, \bibinfo{author}{Qu, G.}, \bibinfo{author}{Wierman, A.}, \bibinfo{year}{2021}.
\newblock \bibinfo{title}{Perturbation-based regret analysis of predictive control in linear time varying systems}.
\newblock \bibinfo{journal}{Advances in Neural Information Processing Systems} \bibinfo{volume}{34}, \bibinfo{pages}{5174--5185}.
\bibitem[{L{\"o}hr et~al.(2020)L{\"o}hr, Klau{\v{c}}o, Fikar and M{\"o}nnigmann}]{lohr2020machine}
\bibinfo{author}{L{\"o}hr, Y.}, \bibinfo{author}{Klau{\v{c}}o, M.}, \bibinfo{author}{Fikar, M.}, \bibinfo{author}{M{\"o}nnigmann, M.}, \bibinfo{year}{2020}.
\newblock \bibinfo{title}{Machine learning assisted solutions of mixed integer mpc on embedded platforms}.
\newblock \bibinfo{journal}{IFAC-PapersOnLine} \bibinfo{volume}{53}, \bibinfo{pages}{5195--5200}.
\bibitem[{L{\"o}hr et~al.(2019)L{\"o}hr, M{\"o}nnigmann, Klau{\v{c}}o and Kal{\'u}z}]{lohr2019mimicking}
\bibinfo{author}{L{\"o}hr, Y.}, \bibinfo{author}{M{\"o}nnigmann, M.}, \bibinfo{author}{Klau{\v{c}}o, M.}, \bibinfo{author}{Kal{\'u}z, M.}, \bibinfo{year}{2019}.
\newblock \bibinfo{title}{Mimicking predictive control with neural networks in domestic heating systems}, in: \bibinfo{booktitle}{2019 22nd International Conference on Process Control (PC19)}, \bibinfo{organization}{IEEE}. pp. \bibinfo{pages}{19--24}.
\bibitem[{Lovelett et~al.(2020)Lovelett, Dietrich, Lee and Kevrekidis}]{lovelett2020some}
\bibinfo{author}{Lovelett, R.J.}, \bibinfo{author}{Dietrich, F.}, \bibinfo{author}{Lee, S.}, \bibinfo{author}{Kevrekidis, I.G.}, \bibinfo{year}{2020}.
\newblock \bibinfo{title}{Some manifold learning considerations toward explicit model predictive control}.
\newblock \bibinfo{journal}{AIChE Journal} \bibinfo{volume}{66}, \bibinfo{pages}{e16881}.
\bibitem[{Lucia(2014)}]{lucia2015robust}
\bibinfo{author}{Lucia, S.}, \bibinfo{year}{2014}.
\newblock \bibinfo{title}{Robust multi-stage nonlinear model predictive control}.
\bibitem[{Lucia et~al.(2013)Lucia, Finkler and Engell}]{lucia2013multi}
\bibinfo{author}{Lucia, S.}, \bibinfo{author}{Finkler, T.}, \bibinfo{author}{Engell, S.}, \bibinfo{year}{2013}.
\newblock \bibinfo{title}{Multi-stage nonlinear model predictive control applied to a semi-batch polymerization reactor under uncertainty}.
\newblock \bibinfo{journal}{Journal of process control} \bibinfo{volume}{23}, \bibinfo{pages}{1306--1319}.
\bibitem[{Lucia and Karg(2018)}]{lucia2018deep}
\bibinfo{author}{Lucia, S.}, \bibinfo{author}{Karg, B.}, \bibinfo{year}{2018}.
\newblock \bibinfo{title}{A deep learning-based approach to robust nonlinear model predictive control}.
\newblock \bibinfo{journal}{IFAC-PapersOnLine} \bibinfo{volume}{51}, \bibinfo{pages}{511--516}.
\bibitem[{Lucia et~al.(2020)Lucia, Navarro, Karg, Sarnago and Lucia}]{lucia2020deep}
\bibinfo{author}{Lucia, S.}, \bibinfo{author}{Navarro, D.}, \bibinfo{author}{Karg, B.}, \bibinfo{author}{Sarnago, H.}, \bibinfo{author}{Lucia, O.}, \bibinfo{year}{2020}.
\newblock \bibinfo{title}{Deep learning-based model predictive control for resonant power converters}.
\newblock \bibinfo{journal}{IEEE Transactions on Industrial Informatics} \bibinfo{volume}{17}, \bibinfo{pages}{409--420}.
\bibitem[{Mayne et~al.(2011)Mayne, Kerrigan, Van~Wyk and Falugi}]{mayne2011tube}
\bibinfo{author}{Mayne, D.Q.}, \bibinfo{author}{Kerrigan, E.C.}, \bibinfo{author}{Van~Wyk, E.}, \bibinfo{author}{Falugi, P.}, \bibinfo{year}{2011}.
\newblock \bibinfo{title}{Tube-based robust nonlinear model predictive control}.
\newblock \bibinfo{journal}{International journal of robust and nonlinear control} \bibinfo{volume}{21}, \bibinfo{pages}{1341--1353}.
\bibitem[{Mayne et~al.(2005)Mayne, Seron and Rakovi{\'c}}]{mayne2005robust}
\bibinfo{author}{Mayne, D.Q.}, \bibinfo{author}{Seron, M.M.}, \bibinfo{author}{Rakovi{\'c}, S.}, \bibinfo{year}{2005}.
\newblock \bibinfo{title}{Robust model predictive control of constrained linear systems with bounded disturbances}.
\newblock \bibinfo{journal}{Automatica} \bibinfo{volume}{41}, \bibinfo{pages}{219--224}.
\bibitem[{Meng et~al.(2022)Meng, Shen and Karimi}]{meng2022emerging}
\bibinfo{author}{Meng, F.}, \bibinfo{author}{Shen, X.}, \bibinfo{author}{Karimi, H.R.}, \bibinfo{year}{2022}.
\newblock \bibinfo{title}{Emerging methodologies in stability and optimization problems of learning-based nonlinear model predictive control: A survey}.
\newblock \bibinfo{journal}{International Journal of Circuit Theory and Applications} \bibinfo{volume}{50}, \bibinfo{pages}{4146--4170}.
\bibitem[{Mesbah et~al.(2022)Mesbah, Wabersich, Schoellig, Zeilinger, Lucia, Badgwell and Paulson}]{mesbah2022fusion}
\bibinfo{author}{Mesbah, A.}, \bibinfo{author}{Wabersich, K.P.}, \bibinfo{author}{Schoellig, A.P.}, \bibinfo{author}{Zeilinger, M.N.}, \bibinfo{author}{Lucia, S.}, \bibinfo{author}{Badgwell, T.A.}, \bibinfo{author}{Paulson, J.A.}, \bibinfo{year}{2022}.
\newblock \bibinfo{title}{Fusion of machine learning and mpc under uncertainty: What advances are on the horizon?}, in: \bibinfo{booktitle}{2022 American Control Conference (ACC)}, \bibinfo{organization}{IEEE}. pp. \bibinfo{pages}{342--357}.
\bibitem[{Norouzi et~al.(2023)Norouzi, Heidarifar, Borhan, Shahbakhti and Koch}]{norouzi2023integrating}
\bibinfo{author}{Norouzi, A.}, \bibinfo{author}{Heidarifar, H.}, \bibinfo{author}{Borhan, H.}, \bibinfo{author}{Shahbakhti, M.}, \bibinfo{author}{Koch, C.R.}, \bibinfo{year}{2023}.
\newblock \bibinfo{title}{Integrating machine learning and model predictive control for automotive applications: A review and future directions}.
\newblock \bibinfo{journal}{Engineering Applications of Artificial Intelligence} \bibinfo{volume}{120}, \bibinfo{pages}{105878}.
\bibitem[{Nubert et~al.(2020)Nubert, K{\"o}hler, Berenz, Allg{\"o}wer and Trimpe}]{nubert2020safe}
\bibinfo{author}{Nubert, J.}, \bibinfo{author}{K{\"o}hler, J.}, \bibinfo{author}{Berenz, V.}, \bibinfo{author}{Allg{\"o}wer, F.}, \bibinfo{author}{Trimpe, S.}, \bibinfo{year}{2020}.
\newblock \bibinfo{title}{Safe and fast tracking on a robot manipulator: Robust mpc and neural network control}.
\newblock \bibinfo{journal}{IEEE Robotics and Automation Letters} \bibinfo{volume}{5}, \bibinfo{pages}{3050--3057}.
\bibitem[{Nurbayeva et~al.(2022)Nurbayeva, Shintemirov and Rubagotti}]{nurbayeva2022deep}
\bibinfo{author}{Nurbayeva, A.}, \bibinfo{author}{Shintemirov, A.}, \bibinfo{author}{Rubagotti, M.}, \bibinfo{year}{2022}.
\newblock \bibinfo{title}{Deep imitation learning of nonlinear model predictive control laws for safe physical human-robot interaction}.
\newblock \bibinfo{journal}{IEEE Transactions on Industrial Informatics} .
\bibitem[{Ortega and Camacho(1996)}]{ortega1996mobile}
\bibinfo{author}{Ortega, J.G.}, \bibinfo{author}{Camacho, E.}, \bibinfo{year}{1996}.
\newblock \bibinfo{title}{Mobile robot navigation in a partially structured static environment, using neural predictive control}.
\newblock \bibinfo{journal}{Control Engineering Practice} \bibinfo{volume}{4}, \bibinfo{pages}{1669--1679}.
\bibitem[{Parisini et~al.(1998)Parisini, Sanguineti and Zoppoli}]{parisini1998nonlinear}
\bibinfo{author}{Parisini, T.}, \bibinfo{author}{Sanguineti, M.}, \bibinfo{author}{Zoppoli, R.}, \bibinfo{year}{1998}.
\newblock \bibinfo{title}{Nonlinear stabilization by receding-horizon neural regulators}.
\newblock \bibinfo{journal}{International Journal of Control} \bibinfo{volume}{70}, \bibinfo{pages}{341--362}.
\bibitem[{Parisini and Zoppoli(1995)}]{parisini1995receding}
\bibinfo{author}{Parisini, T.}, \bibinfo{author}{Zoppoli, R.}, \bibinfo{year}{1995}.
\newblock \bibinfo{title}{A receding-horizon regulator for nonlinear systems and a neural approximation}.
\newblock \bibinfo{journal}{Automatica} \bibinfo{volume}{31}, \bibinfo{pages}{1443--1451}.
\bibitem[{Park et~al.(2023)Park, Babaei, Munoz, Venkat and Hedengren}]{park2023simultaneous}
\bibinfo{author}{Park, J.}, \bibinfo{author}{Babaei, M.R.}, \bibinfo{author}{Munoz, S.A.}, \bibinfo{author}{Venkat, A.N.}, \bibinfo{author}{Hedengren, J.D.}, \bibinfo{year}{2023}.
\newblock \bibinfo{title}{Simultaneous multistep transformer architecture for model predictive control}.
\newblock \bibinfo{journal}{Computers \& Chemical Engineering} , \bibinfo{pages}{108396}.
\bibitem[{Paulson and Mesbah(2020)}]{paulson2020approximate}
\bibinfo{author}{Paulson, J.A.}, \bibinfo{author}{Mesbah, A.}, \bibinfo{year}{2020}.
\newblock \bibinfo{title}{Approximate closed-loop robust model predictive control with guaranteed stability and constraint satisfaction}.
\newblock \bibinfo{journal}{IEEE Control Systems Letters} \bibinfo{volume}{4}, \bibinfo{pages}{719--724}.
\bibitem[{Pin et~al.(2010)Pin, Filippo, Pellegrino, Fenu and Parisini}]{pin2010approximate}
\bibinfo{author}{Pin, G.}, \bibinfo{author}{Filippo, M.}, \bibinfo{author}{Pellegrino, F.A.}, \bibinfo{author}{Fenu, G.}, \bibinfo{author}{Parisini, T.}, \bibinfo{year}{2010}.
\newblock \bibinfo{title}{Approximate off-line receding horizon control of constrained nonlinear discrete-time systems: Smooth approximation of the control law}, in: \bibinfo{booktitle}{Proceedings of the 2010 American Control Conference}, \bibinfo{organization}{IEEE}. pp. \bibinfo{pages}{6268--6273}.
\bibitem[{Pin et~al.(2013)Pin, Filippo, Pellegrino, Fenu and Parisini}]{pin2013approximate}
\bibinfo{author}{Pin, G.}, \bibinfo{author}{Filippo, M.}, \bibinfo{author}{Pellegrino, F.A.}, \bibinfo{author}{Fenu, G.}, \bibinfo{author}{Parisini, T.}, \bibinfo{year}{2013}.
\newblock \bibinfo{title}{Approximate model predictive control laws for constrained nonlinear discrete-time systems: analysis and offline design}.
\newblock \bibinfo{journal}{International Journal of Control} \bibinfo{volume}{86}, \bibinfo{pages}{804--820}.
\bibitem[{Rakovic and Levine(2018)}]{rakovic2018handbook}
\bibinfo{author}{Rakovic, S.V.}, \bibinfo{author}{Levine, W.S.}, \bibinfo{year}{2018}.
\newblock \bibinfo{title}{Handbook of model predictive control}.
\newblock \bibinfo{publisher}{Springer}.
\bibitem[{Ren et~al.(2022)Ren, Alhajeri, Luo, Chen, Abdullah, Wu and Christofides}]{ren2022tutorial}
\bibinfo{author}{Ren, Y.M.}, \bibinfo{author}{Alhajeri, M.S.}, \bibinfo{author}{Luo, J.}, \bibinfo{author}{Chen, S.}, \bibinfo{author}{Abdullah, F.}, \bibinfo{author}{Wu, Z.}, \bibinfo{author}{Christofides, P.D.}, \bibinfo{year}{2022}.
\newblock \bibinfo{title}{A tutorial review of neural network modeling approaches for model predictive control}.
\newblock \bibinfo{journal}{Computers \& Chemical Engineering} , \bibinfo{pages}{107956}.
\bibitem[{Schwenzer et~al.(2021)Schwenzer, Ay, Bergs and Abel}]{schwenzer2021review}
\bibinfo{author}{Schwenzer, M.}, \bibinfo{author}{Ay, M.}, \bibinfo{author}{Bergs, T.}, \bibinfo{author}{Abel, D.}, \bibinfo{year}{2021}.
\newblock \bibinfo{title}{Review on model predictive control: An engineering perspective}.
\newblock \bibinfo{journal}{The International Journal of Advanced Manufacturing Technology} \bibinfo{volume}{117}, \bibinfo{pages}{1327--1349}.
\bibitem[{Scokaert and Mayne(1998)}]{scokaert1998min}
\bibinfo{author}{Scokaert, P.O.}, \bibinfo{author}{Mayne, D.Q.}, \bibinfo{year}{1998}.
\newblock \bibinfo{title}{Min-max feedback model predictive control for constrained linear systems}.
\newblock \bibinfo{journal}{IEEE Transactions on Automatic control} \bibinfo{volume}{43}, \bibinfo{pages}{1136--1142}.
\bibitem[{Shi(2023)}]{shi2023reliable}
\bibinfo{author}{Shi, G.}, \bibinfo{year}{2023}.
\newblock \bibinfo{title}{Reliable Learning and Control in Dynamic Environments: Towards Unified Theory and Learned Robotic Agility}.
\newblock Ph.D. thesis. California Institute of Technology.
\bibitem[{Shi et~al.(2019)Shi, Shi, O’Connell, Yu, Azizzadenesheli, Anandkumar, Yue and Chung}]{shi2019neural}
\bibinfo{author}{Shi, G.}, \bibinfo{author}{Shi, X.}, \bibinfo{author}{O’Connell, M.}, \bibinfo{author}{Yu, R.}, \bibinfo{author}{Azizzadenesheli, K.}, \bibinfo{author}{Anandkumar, A.}, \bibinfo{author}{Yue, Y.}, \bibinfo{author}{Chung, S.J.}, \bibinfo{year}{2019}.
\newblock \bibinfo{title}{Neural lander: Stable drone landing control using learned dynamics}, in: \bibinfo{booktitle}{2019 international conference on robotics and automation (icra)}, \bibinfo{organization}{IEEE}. pp. \bibinfo{pages}{9784--9790}.
\bibitem[{Stellato et~al.(2020)Stellato, Banjac, Goulart, Bemporad and Boyd}]{osqp}
\bibinfo{author}{Stellato, B.}, \bibinfo{author}{Banjac, G.}, \bibinfo{author}{Goulart, P.}, \bibinfo{author}{Bemporad, A.}, \bibinfo{author}{Boyd, S.}, \bibinfo{year}{2020}.
\newblock \bibinfo{title}{{OSQP}: an operator splitting solver for quadratic programs}.
\newblock \bibinfo{journal}{Mathematical Programming Computation} \bibinfo{volume}{12}, \bibinfo{pages}{637--672}.
\newblock \URLprefix \url{https://doi.org/10.1007/s12532-020-00179-2}, \DOIprefix\doi{10.1007/s12532-020-00179-2}.
\bibitem[{Tank and Hopfield(1986)}]{tank1986simple}
\bibinfo{author}{Tank, D.}, \bibinfo{author}{Hopfield, J.}, \bibinfo{year}{1986}.
\newblock \bibinfo{title}{Simple'neural'optimization networks: An a/d converter, signal decision circuit, and a linear programming circuit}.
\newblock \bibinfo{journal}{IEEE transactions on circuits and systems} \bibinfo{volume}{33}, \bibinfo{pages}{533--541}.
\bibitem[{Vaupel et~al.(2019)Vaupel, Caspari, Hamacher, Huster, Mhamdi, Kevrekidis and Mitsos}]{vaupel2019artificial}
\bibinfo{author}{Vaupel, Y.}, \bibinfo{author}{Caspari, A.}, \bibinfo{author}{Hamacher, N.C.}, \bibinfo{author}{Huster, W.R.}, \bibinfo{author}{Mhamdi, A.}, \bibinfo{author}{Kevrekidis, I.G.}, \bibinfo{author}{Mitsos, A.}, \bibinfo{year}{2019}.
\newblock \bibinfo{title}{Artificial neural networks for real-time model predictive control of organic rankine cycles for waste heat recovery}, in: \bibinfo{booktitle}{Proceedings of the 5th international seminar on ORC power systems}, pp. \bibinfo{pages}{1--8}.
\bibitem[{Vaupel et~al.(2020)Vaupel, Hamacher, Caspari, Mhamdi, Kevrekidis and Mitsos}]{vaupel2020accelerating}
\bibinfo{author}{Vaupel, Y.}, \bibinfo{author}{Hamacher, N.C.}, \bibinfo{author}{Caspari, A.}, \bibinfo{author}{Mhamdi, A.}, \bibinfo{author}{Kevrekidis, I.G.}, \bibinfo{author}{Mitsos, A.}, \bibinfo{year}{2020}.
\newblock \bibinfo{title}{Accelerating nonlinear model predictive control through machine learning}.
\newblock \bibinfo{journal}{Journal of process control} \bibinfo{volume}{92}, \bibinfo{pages}{261--270}.
\bibitem[{Von~Luxburg and Sch{\"o}lkopf(2011)}]{von2011statistical}
\bibinfo{author}{Von~Luxburg, U.}, \bibinfo{author}{Sch{\"o}lkopf, B.}, \bibinfo{year}{2011}.
\newblock \bibinfo{title}{Statistical learning theory: Models, concepts, and results}, in: \bibinfo{booktitle}{Handbook of the History of Logic}. \bibinfo{publisher}{Elsevier}. volume~\bibinfo{volume}{10}, pp. \bibinfo{pages}{651--706}.
\bibitem[{Wang et~al.(2022)Wang, Li and Xu}]{wang2022learning}
\bibinfo{author}{Wang, R.}, \bibinfo{author}{Li, H.}, \bibinfo{author}{Xu, D.}, \bibinfo{year}{2022}.
\newblock \bibinfo{title}{Learning model predictive control law for nonlinear systems}, in: \bibinfo{booktitle}{2022 5th International Symposium on Autonomous Systems (ISAS)}, \bibinfo{organization}{IEEE}. pp. \bibinfo{pages}{1--6}.
\bibitem[{Wen et~al.(2009)Wen, Lan and Shih}]{wen2009review}
\bibinfo{author}{Wen, U.P.}, \bibinfo{author}{Lan, K.M.}, \bibinfo{author}{Shih, H.S.}, \bibinfo{year}{2009}.
\newblock \bibinfo{title}{A review of hopfield neural networks for solving mathematical programming problems}.
\newblock \bibinfo{journal}{European Journal of Operational Research} \bibinfo{volume}{198}, \bibinfo{pages}{675--687}.
\bibitem[{Witsenhausen(1968)}]{witsenhausen1968minimax}
\bibinfo{author}{Witsenhausen, H.}, \bibinfo{year}{1968}.
\newblock \bibinfo{title}{A minimax control problem for sampled linear systems}.
\newblock \bibinfo{journal}{IEEE Transactions on Automatic Control} \bibinfo{volume}{13}, \bibinfo{pages}{5--21}.
\bibitem[{Zhang et~al.(2019)Zhang, Bujarbaruah and Borrelli}]{zhang2019safe}
\bibinfo{author}{Zhang, X.}, \bibinfo{author}{Bujarbaruah, M.}, \bibinfo{author}{Borrelli, F.}, \bibinfo{year}{2019}.
\newblock \bibinfo{title}{Safe and near-optimal policy learning for model predictive control using primal-dual neural networks}, in: \bibinfo{booktitle}{2019 American Control Conference (ACC)}, \bibinfo{organization}{IEEE}. pp. \bibinfo{pages}{354--359}.
\bibitem[{Zhang et~al.(2020)Zhang, Bujarbaruah and Borrelli}]{zhang2020near}
\bibinfo{author}{Zhang, X.}, \bibinfo{author}{Bujarbaruah, M.}, \bibinfo{author}{Borrelli, F.}, \bibinfo{year}{2020}.
\newblock \bibinfo{title}{Near-optimal rapid mpc using neural networks: A primal-dual policy learning framework}.
\newblock \bibinfo{journal}{IEEE Transactions on Control Systems Technology} \bibinfo{volume}{29}, \bibinfo{pages}{2102--2114}.

\end{thebibliography}

\end{document}